\documentclass{emulateapj}
\usepackage{amssymb,amsmath,graphicx,longtable,subfigure,wrapfig}
\usepackage{rotating}
\usepackage[colorlinks,linkcolor=blue,anchorcolor=green,citecolor=blue]{hyperref}
\usepackage{natbib}

\shorttitle{GMRT study of head$-$tail galaxies}
\shortauthors{Sebastian, Lal \& Rao}

\begin{document}

\def\func#1{\mathop{\rm #1}\nolimits}
\def\unit#1{\mathord{\thinspace\rm #1}}


\title{Giant Metrewave Radio Telescope Observations of Head$-$Tail Radio Galaxies}

%% anywhere in the paper, not just in the front matter.2

\author{Biny Sebastian, Dharam V. Lal and A. Pramesh Rao}
\affil{National Center for Radio Astrophysics - Tata Institute of Fundamental Research
               Post Box 3, Ganeshkhind P.O., Pune 41007, India; biny@ncra.tifr.res.in}
%\altaffiltext{National Center for Radio Astrophysics - Tata Institute of Fundamental Research
%               Post Box 3, Ganeshkhind P.O., Pune 41007, India; biny@ncra.tifr.res.in}
%\email{biny@ncra.tifr.res.in}

\begin{abstract}

We present results from a study of seven large known head$-$tail radio galaxies
based on observations using the Giant Metrewave Radio Telescope at 240 and 610
MHz.
These observations are used to study the radio morphologies and distribution of the spectral indices
across the sources. The overall morphology of the radio tails of these sources
is suggestive of random motions of the optical host around the cluster
potential. The presence of the multiple bends and wiggles in several head$-$tail
sources is possibly due to the precessing radio jets. We find steepening of
the spectral index along the radio tails. The prevailing equipartition magnetic
field also decreases along the radio tails of these sources. These steepening
trends are attributed to the synchrotron aging of plasma toward the ends of
the tails.  The dynamical ages of these sample sources have been estimated to
be $\sim$10$^8$~yr, which is a factor of six more than the age estimates from
the radiative losses due to synchrotron cooling.
%We also calculated the average spectral ages assuming an
%equipartition magnetic field and a comparison between these two estimates of
%ages are done.

\end{abstract}





\section{Introduction}
\label{intro}

It is well known that many radio galaxies have a double radio structure, with
the two extended components symmetrically located with respect to the parent
host galaxy.
%In head$-$tail radio galaxies 
Head$-$tail galaxies discovered by \cite{1968MNRAS.138....1R}
occur in clusters of galaxies and are characterized by a head identified
with the optical galaxy and two tails sweeping back from the head %(Figure~\ref{all-ht})
forming an angle with the galaxy at the apex.
 These sources are
understood to be Fanaroff$-$Riley type~I \citep{1974MNRAS.167P..31F} radio sources
moving through the gas in
the cluster, and the shape of the source is due to the diffuse radio$-$emitting
plasma being decelerated by the intracluster medium
\citep[ICM,][]{1972Natur.237..269M, 1998MNRAS.301..609B}. \cite{1973A&A....26..423J} suggested two models based on this idea:
the independent blob model and the magnetospheric model.
According to the independent
blob model, blobs of plasma were ejected from the galaxy, and they trail behind the
galaxy because of the ram pressure that acts on them.
The magnetospheric model has strong magnetic fields in the
galaxy that are stretched out by the supersonic motion of the galaxy.
The latter model sought to explain the morphology by using the motion of relativistic
electrons ejected from the galaxy's nuclear region along these  magnetic field
lines. 
%Briefly, the difficulties of the former were considered in their latter model and the polarisation data on
%NGC\,1265 \citep{1973A&A....26..413M} was found to be consistent with the latter model.
However, later ram pressure bending models were preferred  over the
magnetospheric model, and, subsequently, the beam models that involved the bending of
continuous jets by the ram pressure replaced independent blob models
\citep{1979Natur.279..770B,1979ApJ...234..818J}, which is supported by
several three-dimensional hydrodynamic simulations
\citep{1984Natur.310...33W, 1992ApJ...393..631B, 2011ApJ...730...22P}.

Additionally, the role played by magnetic fields in the evolution and dynamics of head$-$tail
galaxies is barely explored.
\cite{1973A&A....26..413M} noticed that
the polarization distributions of head$-$tail galaxies become
more ordered and also the polarization percentage increases as one moves along
the tail away from the host galaxy.
This increase in the polarization fraction was one of the
reasons that prompted \cite{1973A&A....26..423J} to propose a magnetospheric
model. The increase in polarization percentage is higher than what is expected according to
synchrotron theory, due to the steepening of the spectrum, and the polarization
vectors were noticed to be aligned across the tail perpendicular to the ridge
line pointing toward a very ordered magnetic field that runs parallel to the
tail \citep{1998A&A...331..475F}.

Recently, efforts to reproduce the observed morphology of tailed sources
using three-dimensional magnetohydrodynamic simulations
are being undertaken \citep{2017PhPl...24d1402J,2017ApJ...839...14G}. Head$-$tail radio sources are also used to identify clusters and also as probes of cluster properties
\citep{2000ApJ...531..118B, 2001AJ....121.2915B, 2003AJ....125.1635B, 2011AJ....141...88W, 2016arXiv161100746P}.

\begin{table*}[tbph]
        \centering
        \caption{Sources. Column~1: name as used in this paper;
the name of the cluster is given below it in parentheses.
Column~2: redshift. Column~3: right ascension. Column~4: declination.
Column~5: angular size . Column~6: linear scale for our adopted
cosmology. Column~7: date of observation.  Column~8: bandwidth.  Column~9: center frequency.
Column~10:flux density calibrator. Column~11: phase calibrator. Column~12: on-source integration
time.}
        \label{obs-log}
 \begin{tabular}{lccccccccccc}
\hline \hline
Object          & $z$ & R.A.          &  Decl.       & Size  & Linear scale & Obs. date  & $\Delta\nu$  & $\nu_{\rm cen.}$ & Flux & Phase & Int. time \\
                &        &             &             &($\prime$)& \multicolumn{2}{l}{(kpc~arcsec$^{-1}$)}  &  \multicolumn{2}{c}{(MHz)} & \multicolumn{2}{c}{calibrators} & (hr) \\
(1) & (2) & (3) & (4) & (5) & (6) & (7) & (8) & (9) & (10) & (11) & (12) \\
\hline
PKS\,B0053$-$016& 0.043 & 00\,56\,02.91 &$-$01\,20\,04.5&  5.0 & 0.841  & 2016 Jun 29 & 32 & 322.6/ & 3C48   & 0025$-$260  & 2\\
(Abell\,119)    &       &             &             &      &   			& 2016 Jul 05 & 32 & 608.0  & 3C48   & 0025$-$260  & 2.4\\
PKS\,B0053$-$015& 0.038 & 00\,56\,25.70 &$-$01\,15\,46.8&  6.0 & 0.748  & 2016 Jun 29 & 32 & 322.6/ & 3C48   & 0025$-$260  & 2\\
(Abell\,119)    &       &   \,  \,      &     \,  \,    &      &   		& 2016 Jul 05 & 32 & 608.0  & 3C48   & 0025$-$260  & 2.4\\
IC\,310         & 0.019 & 03\,16\,42.77 &$+$41\,19\,29.6&  8.5 & 0.385  & 2002 Dec 21 & ~8 & 240.3/ & 3C48,  & 0314$+$432, & 6.5\\
(Abell\,426)    &       &   \,  \,      &     \,  \,    &      &   &            &    & 609.6  & 3C147  & 0348$+$338  & 6.5\\
NGC\,1265       & 0.025 & 03\,18\,14.86 &$+$41\,51\,27.6& 10.5 & 0.502 	& 2002 Dec 20 & ~8 & 240.3/ & 3C48,  & 0348$+$338  & 6\\
(Abell\,426)    &       &   \,  \,      &     \,  \,    &      &   &            &    & 609.6  & 3C286  &             & 6\\
GB6\,B0335$+$096& 0.038 & 03\,38\,14.09 &$+$10\,05\,03.9&  8.5 & 0.748 	& 2009 Dec 23 & 16 & 618.3/ & 3C48   & 0521$+$166  & 3.5\\
(2A\,0335$+$096)&       &   \,  \,      &     \,  \,    &      &   		& 2002 May 05 & ~8 & 240.3  & 3C48,  & 0323$+$055, & 1.25\\
                &       &   \,  \,      &     \,  \,    &      &   &            &    &        & 3C147  & 0025$-$260  & \\
IC\,711         & 0.032 & 11\,34\,45.66 &$+$48\,57\,21.5& 15.0 & 0.636 	& 2002 Dec 23 & ~8 & 240.3/ & 3C147, & 1219$+$484, & 6\\
(Abell\,1314)   &       &   \,  \,      &     \,  \,    &      &   &            &    & 609.6  & 3C286  & 1252$+$565  & 6\\
NGC\,7385       & 0.026 & 22\,49\,54.57 &$+$11\,36\,30.1& 12.0 &  0.522 & 2002 Dec 23 & ~8 & 240.3/ & 3C286, & 2232$+$117, & 5.5\\
(Cul\,2247$+$113)&       &             &             &     &    &            &    & 609.6  & 3C48  & 2250$+$143  & 5.5\\
%4C\,13.17a      & 0.025 &             &             &  7.0&    &            &    &        &        & & \\
%3C\,264         & 0.038 &             &             &  9.0&    &            &    &        &        & & \\
%NGC\,6109       & 0.032 &             &             & 12.0&    &            &    &        &        & & \\
%B1709$+$397     & 0.026 &             &             &  5.5&    &            &    &        &        & & \\
\hline
 \end{tabular}
 \end{table*}

\cite{1976ApJ...203..313P} studied the spatial variation of the spectral index, flux density, and polarization of a few  head$-$tail galaxies.
They find that head$-$tail radio galaxy evolves into three distinct regions.
The first region, which is the
closest to the core, has a brightness peak from which it starts to drop along
with an increase in the spectral index. This was thought to be due to particle
acceleration, expansion, and synchrotron losses that occur in this region.
The second region has a roughly constant luminosity and spectral index, which was
attributed to the competition between the acceleration of particles due to
turbulence and the synchrotron losses. According to ram pressure bending
models, a galaxy under the influence of the central cluster potential must be
moving at supersonic velocities. A bow shock is expected to form at the leading
edge of the optical galaxy, causing this turbulence farther downstream.
Finally, the
third region is where the synchrotron losses dominate, and a steady increase in
the spectral index along with a decrease in flux density is observed.

Here, we present a low-frequency study of a sample of seven head$-$tail radio
galaxies using the Giant Metrewave Radio Telescope (GMRT). The radio
morphologies and spatial structure in the radio spectrum are studied in order
to understand the trends discussed above. We also present the variation of
equipartition parameters, including magnetic field, along the head$-$tail radio
galaxy.  In a subsequent paper, we will discuss the total
intensity and brightness gradient, the X-ray properties of these head$-$tail
radio sources, and the morphological properties of a propagating jet as it
crosses its interstellar medium (ISM) into an ICM.

This paper is organized as follows,
Sections~\ref{ht-sample} and \ref{gmrt-obs}, respectively, describe the
head$-$tail radio galaxy sample and GMRT observations along with data analyses.
Radio images and spectral index maps of the head$-$tail radio galaxies, giving a
brief description of the morphological and spectral features
and physical parameters, are presented in Section~\ref{gmrt-result}.
%Radio morphologies and spectra are presented in Section~\ref{gmrt-result}.
We present a discussion from our observations in light of our understanding
from the literature for the sample of head$-$tail radio galaxies in Section~\ref{ht-discuss}
and summarize the salient conclusions of our study in Section~\ref{ht-concl}.
We define radio spectral index, $\alpha$, via $S_\nu \propto \nu^\alpha$, where
$S_\nu$ and $\nu$ are flux density and frequency, respectively.
We adopt a cosmology
with $H_0$ = 69.6 km s$^{-1}$ Mpc$^{-1}$, $\Omega_M$ = 0.286,
and $\Omega_\Lambda$ = 0.714 \citep{spergel07}.

\section{Head$-$tail Galaxy Sample}
\label{ht-sample}

\cite{1997ApJS..108...41O} have surveyed over 500 Abell clusters of galaxies using the
Very Large Array (VLA) at $\lambda$= 21 cm, and they presented radio maps of sources with complex or
multiple features.
They detect and list candidate head$-$tail radio sources in these clusters.
From this list of candidate head$-$tail radio sources, along with the head$-$tail
sources known in the literature, we choose only nearby, $z$ $<$ 0.05 and
confirmed head$-$tail galaxies
with an angular size of galaxy tail larger than 5 arcmin as our sample source.
These 11 sources were observed with the GMRT
during cycles 03 and 04 (03DVL02, 04DVL01).
The data for four sources were badly affected by GMRT hardware correlator issues, so we do not include them here.
Our final sample consisted of seven head$-$tail radio galaxies, listed in Table~\ref{obs-log}.
%(1) source name along with name of the cluster (in parentheses);
%(2) redshift;
%(3, 4) right ascension (R.A.) and declination (dec.);
%(5) angular size;
%(6) linear scale for our adopted cosmology;
%(7) date of observation;
%(8, 9) bandwidth and center observing frequency;
%(10, 11) corresponding flux density and phase calibration sources; and
%(12) on-source integration time.

\section{GMRT observations}
\label{gmrt-obs}

Our sample of seven head$-$tail radio galaxies was observed using the
simultaneous dual$-$frequency(240 MHz and 610 MHz) mode of GMRT, except for two
sources, where data at 240 MHz were marred by radio frequency
interference (RFI) and were later observed at 325 MHz.
Table~\ref{obs-log} gives the details of the observations. 
The columns are as follows:
(1) source name along with the name of the cluster (in parentheses);
(2) redshift;
(3, 4) R.A. and decl.;
(5) angular size (arcmin);
(6) linear scale (kpc~arcsec$^{-1}$) for our adopted cosmology;
(7) date of observation;
(8, 9) bandwidth and center observing frequency;
(10, 11) corresponding flux density and phase calibration sources; and
(12) on-source integration time.

The GMRT is a low-frequency interferometer
with a hybrid configuration. One-half of the antennas out of the total of 30 antennas are
located within a central region of size 1.1 km, and the other half are
distributed along an approximate `Y' shape with a maximum baseline length of 
$\sim$25~km. The presence of a large number of antennas in the central square,
along with enhanced surface brightness at lower frequencies, makes GMRT an ideal
instrument to obtain high-resolution images of the head$-$tail galaxies without
missing much information at extended scales \citep{2007MNRAS.374.1085L}. The
angular resolution of these observations corresponds to at least 20 synthesized
beams across the source, which is sufficient to provide a detailed radio
morphology and allow the study of spectral structure in these sources.

\subsection{Data reduction}

Data sets for several target sources were flagged and calibrated using
the \textsc{flagcal} package \citep{2012ExA....33..157P},
and imaged using the classic package(`\textsc{aips}'),
except for PKS\,B0053$-$016, PKS\,B0053$-$015 and GB6 B0335$+$0955, which were
reduced using standard data-reduction procedures in classic `\textsc{aips}'.
Flux density calibrators observed at the beginning and the end
of the observations were used to calibrate the complex gains of the visibilities
and correct for the bandpass shape.
Phase calibration sources observed typically once every 35 minutes
were used to correct for variations in the amplitude and phase across time.
Spectral channels affected by RFI  were
flagged, and the calibrated data were averaged in frequency using \textsc{aips} task `\textsc{splat}'
before imaging to reduce the data volume.
Care was taken to avoid bandwidth smearing.

\begin{figure*}
 \centering
 \includegraphics[width=23cm,angle=90]{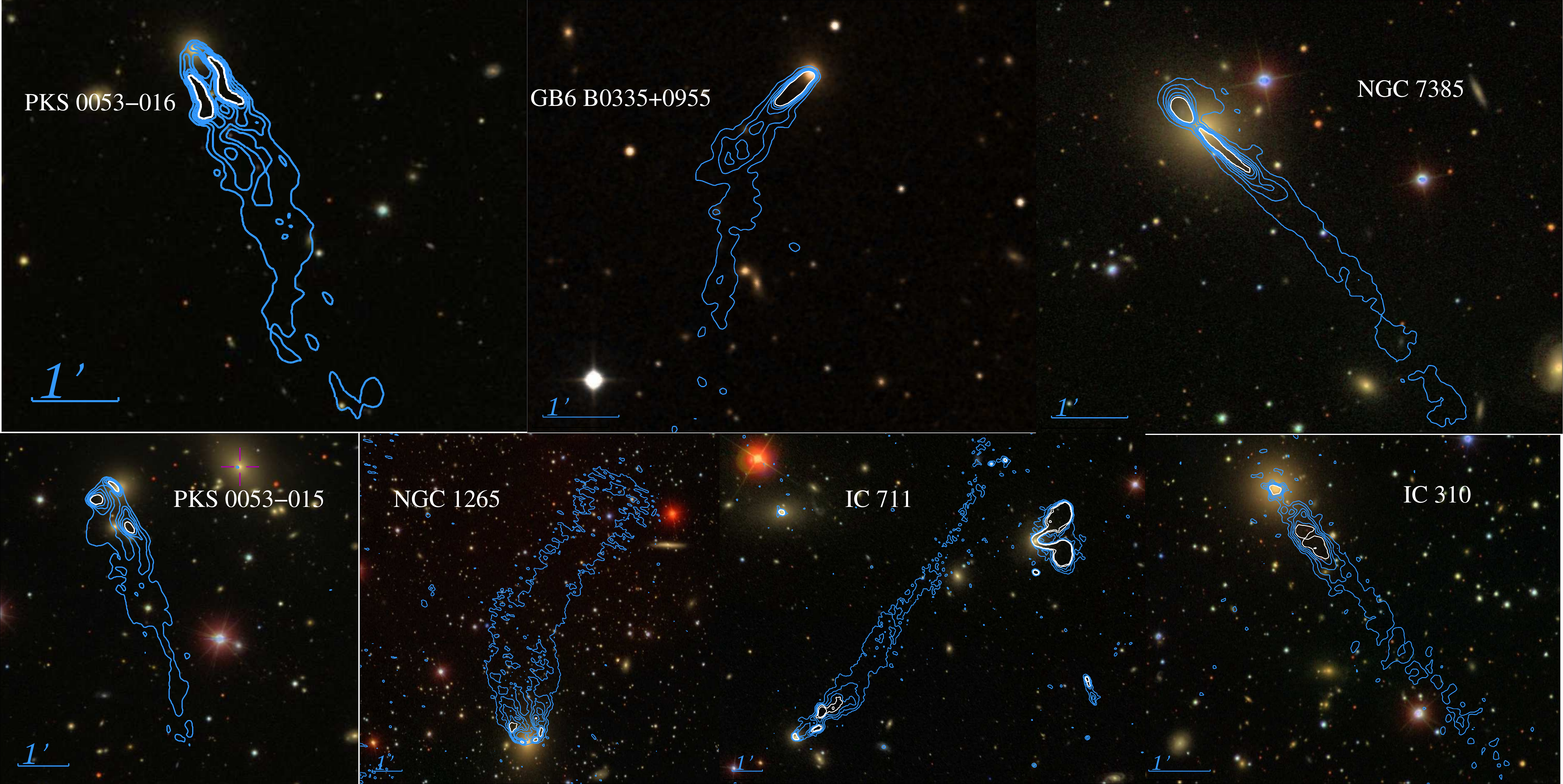}
 \caption{High-resolution, high sensitivity GMRT radio images (contour maps)
at 610 MHz overlaid on the SDSS $gri$-color composite images of our sample sources.
In clockwise order from top left, the sources are PKS\,0053$-$016,
GB6\,B0335$+$0955, NGC\,7385, IC\,310, IC\,711, NGC\,1265 and PKS\,0053$-$015.}
 \label{all-ht}
\end{figure*}

The wide field of view was divided into a number of overlapping facets
for imaging.
After three to four rounds of phase-only self-calibration,
with typical solution interval timescales of 2~minutes for self-calibration, a final amplitude and phase self-calibration were done.  These sets of facets were interpolated to a single image
using \textsc{aips} task `\textsc{flatn}'
centered on the pointing position and were corrected for the primary beam using the \textsc{aips} task
`\textsc{pbcor}'.
Figure~\ref{all-ht} presents high-resolution, high-sensitivity radio images
at 610 MHz (contour maps) overlaid on the SDSS~$gri$-color composite images of all seven
head$-$tail radio galaxies (see Table~\ref{obs-log}).
To enhance the detailed source structure with a reasonably good sensitivity,
both high and low angular resolution images at both 240/325 MHz and 610 MHz frequencies
were constructed.
The default high-angular-resolution images were made with
$\sim$5$^{\prime\prime}$ and $\sim$10$^{\prime\prime}$ synthesized beams,
and the low-angular-resolution images were made by tapering the
visibility data at 15 k$\lambda$  and restored using either a 15$^{\prime\prime}$
or a 20$^{\prime\prime}$ synthesized beam.
The misalignment between the matched resolution images at the two frequencies, if any,
was corrected, and the spectral index maps were constructed using the \textsc{aips}
task `\textsc{comb}'.  We blanked pixels where the total intensity
level was less than three times the rms noise level (see also Sec.~\ref{rad_morph})
at a given frequency band in order to make spectral index maps.

\section{Results}
\label{gmrt-result}

\subsection{Radio morphology}
\label{rad_morph}

The radio maps of our head$-$tail sample radio galaxies at frequencies 610 MHz
and 240/325 MHz are shown in Figures~\ref{0053-016}--\ref{n7385} with angular
resolutions of $\sim$5$^{\prime\prime}$ and $\sim$10$^{\prime\prime}$, respectively. The maps are ordered in R.A.  
Similar to earlier published radio continuum images
of X-shaped and FR\,II radio galaxies \citep[e.g.,][etc.]{2005ASPC..345..289L,2007MNRAS.374.1085L,2008MNRAS.390.1105L} the dynamic ranges in
the maps are 900$-$5000 and the local noise in the vicinity of our
targets are occasionally higher than the noise in empty regions.  The contour
levels shown in these figures are based on the rms noise in the vicinity of our
targets, which is generally more, with the first contour level being 3--5 times
this rms noise level.  The upper panel shows the default high-resolution
images, while the lower panel shows the low-resolution images.
The integrated flux densities of all the sources
along with previous measurements from
the literature are plotted in Figure~\ref{int-flux}.
Our estimates at both frequencies, 240/325 MHz
and 610 MHz, agree well with other measurements.
A brief
description of the radio morphologies at these observing frequencies and angular resolutions
of our sample sources is discussed below.

\subsubsection{PKS\,B0053$-$016}

\cite{1998A&A...331..475F} and \cite{1985AJ.....90..927O} have studied
PKS\,B0053$-$016 and PKS\,0053$-$015 high-frequency using VLA.
Also from their polarization studies, \cite{1993A&A...280...63M} find magnetic
field components aligned with the tail.  These two galaxies are members of the
ACO\,119 cluster.
%The morphology of PKS\,0053$-$015 is described in next section.
Figure~\ref{all-ht} suggests that the host galaxy of PKS\,0053$-$015
%The morphology of PKS\,0053$-$015 (Figure~\ref{all-ht}) suggests that the host galaxy
is moving toward
the direction of the cluster center, with two very symmetric tails trailing
behind. The overall motion of the galaxy follows a straight line without much
evidence for a curved trajectory around the cluster center.

\begin{figure*}
\begin{center}
\begin{tabular}{c}
\includegraphics[width=18cm]{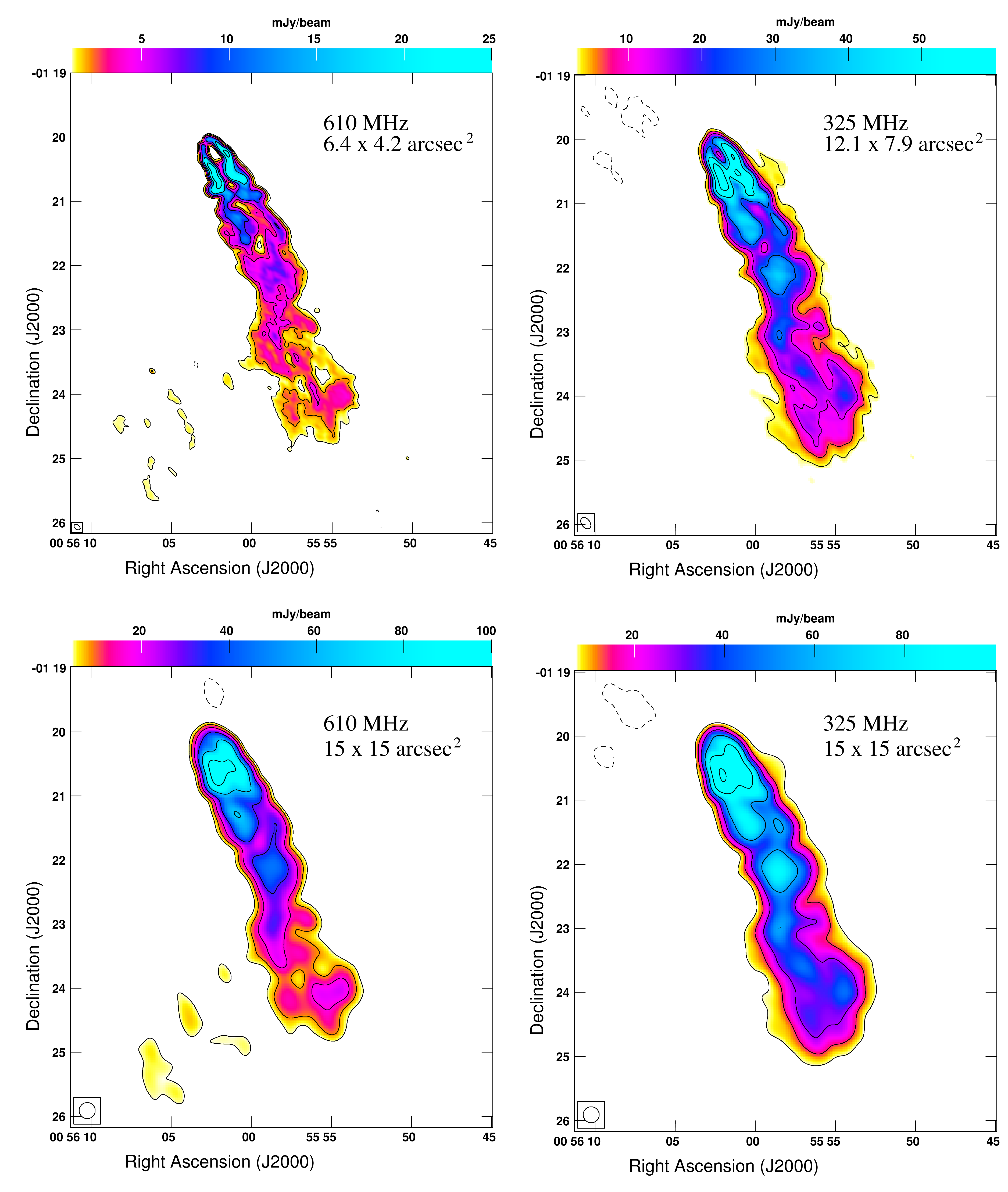} 
\end{tabular}
\caption{Full synthesis images of PKS\,B0053$-$016. Top left panel:
image at 610 MHz with a beam size of
%6.4$^{\prime\prime}$ $\times$ 4.2$^{\prime\prime}$ at a PA of 52.1$^\circ$,
$6\hbox{$.\!\!^{\prime\prime}$}4$ $\times$ $4\hbox{$.\!\!^{\prime\prime}$}2$
at a PA $52\hbox{$.\!\!^{\circ}$}1$,
the peak surface brightness is 51.7 mJy~beam$^{-1}$,
the error bar at a source-free location is 0.3 mJy~beam$^{-1}$ and
the contour levels are 1.0 $\times$ $-$2, $-$1, 1, 2, 4, 8, 16, 32, 64, 128 mJy~beam$^{-1}$.
Top right panel:
image at 325 MHz with a beam size of
%12.1$^{\prime\prime}$ $\times$ 7.9$^{\prime\prime}$ at a PA of $-$36.7$^\circ$,
$12\hbox{$.\!\!^{\prime\prime}$}1$ $\times$ $7\hbox{$.\!\!^{\prime\prime}$}9$
at a PA $-36\hbox{$.\!\!^{\circ}$}7$,
the peak surface brightness is 151.3 mJy~beam$^{-1}$,
the error bar at a source-free location is 0.8 mJy~beam$^{-1}$ and
the contour levels are  3.2 $\times$ ($-$1, 1, 2, 4, 8, 16, 32, 64) mJy~beam$^{-1}$.
Bottom left panel:
low-resolution image at 610 MHz with
a beam size of
15$^{\prime\prime}$ $\times$ 15$^{\prime\prime}$,
the peak surface brightness is 176.1 mJy~beam$^{-1}$,
the error bar at a source-free location is 1.0 mJy~beam$^{-1}$ and
the contour levels are  4.0 $\times$ ($-$2, $-$1, 1, 2, 4, 8, 16, 32, 64, 128) mJy~beam$^{-1}$.
Bottom right panel:
low-resolution image at 325 MHz with
a beam size of 15$^{\prime\prime}$ $\times$ 15$^{\prime\prime}$,
the peak surface brightness is 236.5 mJy~beam$^{-1}$,
the error bar at a source-free location is 2.0 mJy~beam$^{-1}$ and
the contour levels are  7.0 $\times$ ($-$2, $-$1, 1, 2, 4, 8, 16, 32, 64, 128) mJy~beam$^{-1}$.}
    \label{0053-016}
\end{center}
\end{figure*}

The radio morphology is quite symmetric and has an overall angular extent of
~6$^{\prime}$,  and the core is not very prominent. The high-resolution contour
map (Figure~\ref{0053-016}) at 610 MHz resolves the jets that are
ejected from the core, as two distinct radio jets. Slightly away from the core the intensity increases and
the jet widens. %This is similar to the flaring seen close to the core in FR\,I
radio galaxies. Both the jets also seem to follow a helical trajectory, and
the tails seem to show symmetric wiggles. This substantiates the speculation that this
has something to do with the origin of the jets, and is not due to any interaction
with the ambient medium or instabilities
\citep{Kharb}. This helicity is presumably introduced by the precession of the jets.

\subsubsection{PKS\,B0053$-$015}\label{pksbmorph}     

%PKS\,0053$-$015 is shown in Figure~\ref{0056-015}.
%These galaxies are members of the ACO-119 cluster.
Figure~\ref{all-ht} shows that the host galaxy is not at the leading edge of
the tails, unlike most head$-$tail galaxies. The jets are clearly ejected at wide
angles. The peculiar morphology might be a consequence of the fact that one of
the jets is not ejected at right angles but instead at a smaller angle with the
direction of motion of the galaxy. This might also result from projection
effects or an interplay of both. If the former is the reason, then the bending
of the northern jet, which has a velocity component in the direction of motion,
suggests that the jet experiences a very high ram pressure. The projected
angular distance from the center of the cluster is only 2$^{\prime}$.4. It is
thus possible that the galaxy is close to the cluster center, which would
in turn mean higher galaxy velocity and higher ICM density and thus higher ram
pressure. Galaxies in a cluster are expected to follow an orbital motion around
the cluster potential well.

\begin{figure*}
\begin{center}
\begin{tabular}{c}
\includegraphics[width=18cm]{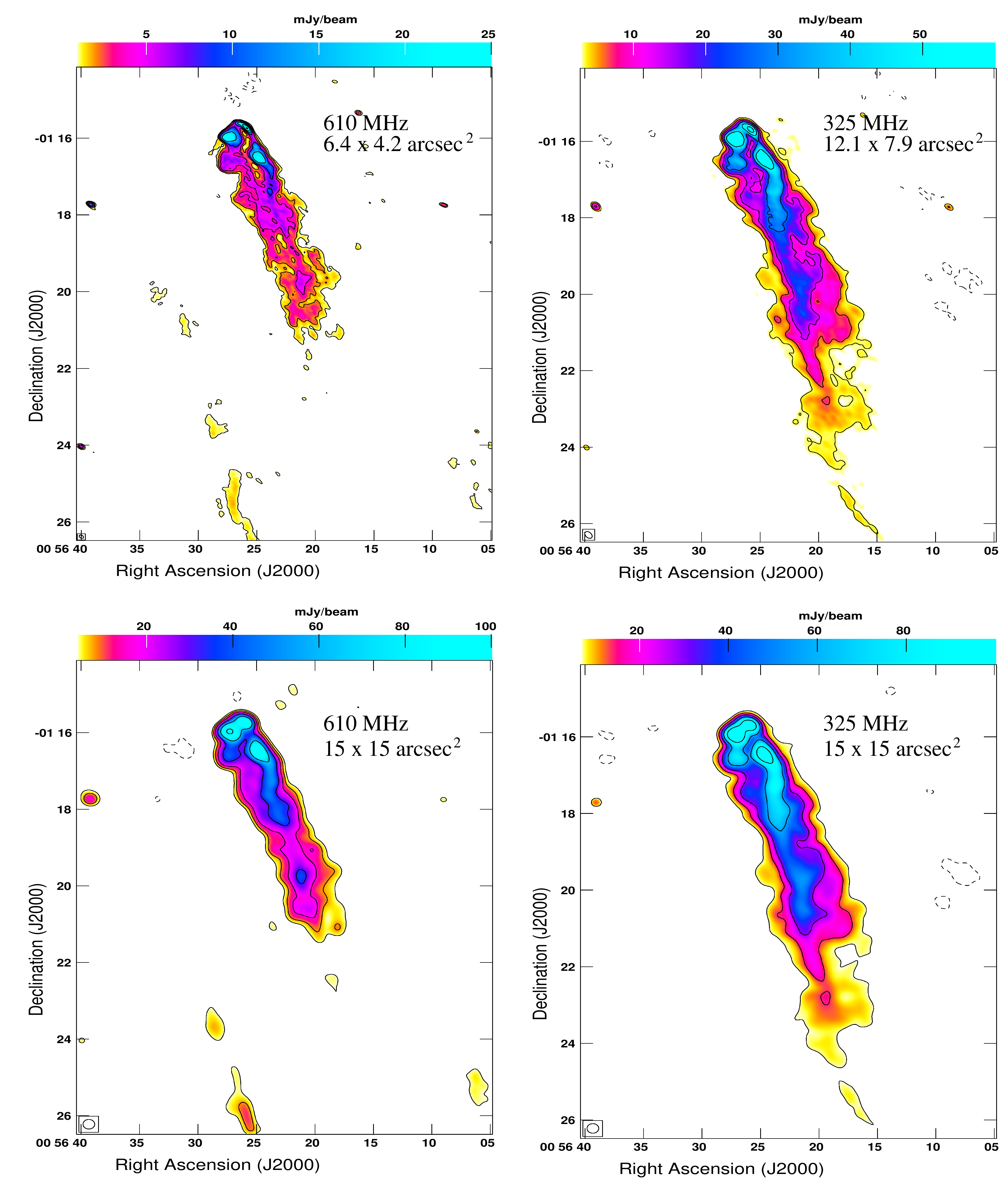}
\end{tabular}
\caption{Full synthesis images of PKS\,B0053$-$015. Top left panel:
image at 610 MHz with a beam size of
%6.4$^{\prime\prime}$ $\times$ 4.2$^{\prime\prime}$ at a PA of 52.1$^\circ$,
$6\hbox{$.\!\!^{\prime\prime}$}4$ $\times$ $4\hbox{$.\!\!^{\prime\prime}$}2$ at a PA $52\hbox{$.\!\!^{\circ}$}1$,
the peak surface brightness is 67.1 mJy~beam$^{-1}$,
the error~bar at a source-free location is 0.3 mJy~beam$^{-1}$ and
the contour levels are  1.0 $\times$ ($-$2, 2, 4, 8, 16, 32, 64, 128) mJy~beam$^{-1}$.
Top right panel:
image at 325 MHz with a beam size of
%12.1$^{\prime\prime}$ $\times$ 7.9$^{\prime\prime}$ at a PA of $-$36.7$^\circ$,
$12\hbox{$.\!\!^{\prime\prime}$}1$ $\times$ $7\hbox{$.\!\!^{\prime\prime}$}9$ at a PA $-36\hbox{$.\!\!^{\circ}$}7$,
the peak surface brightness is  134.9 mJy~beam$^{-1}$,
the error~bar at a source-free location is 0.8 mJy~beam$^{-1}$ and
the contour levels are  3.2 $\times$ ($-$1, 1, 2, 4, 8, 16, 32, 64) mJy~beam$^{-1}$.
Bottom left panel:
low-resolution image at 610 MHz with
a beam size of 15$^{\prime\prime}$ $\times$ 15$^{\prime\prime}$,
the peak surface brightness is 137.8 mJy~beam$^{-1}$,
the error~bar at a source-free location is 1.0 mJy~beam$^{-1}$ and
the contour levels are  4.0 $\times$ ($-$2, $-$1, 1, 2, 4, 8, 16, 32, 64, 128) mJy~beam$^{-1}$.
Bottom right panel:
low-resolution image at 325 MHz with
a beam size of 15$^{\prime\prime}$ $\times$ 15$^{\prime\prime}$,
the peak surface brightness is 196.7 mJy~beam$^{-1}$,
the error~bar at a source-free location is 2.0 mJy~beam$^{-1}$ and
the contour levels are  7.0 $\times$ ($-$2, $-$1, 1, 2, 4, 8, 16, 32, 64, 128) mJy~beam$^{-1}$.}
    \label{0056-015}
\end{center}
\end{figure*}

The radio morphology of PKS\,0053$-$015,
shown in Figure~\ref{0056-015} is very similar to PKS\,B0053$-$016.
The angular size of this head$-$tail galaxy is 5$^{\prime}$.0. The core is bright,
and an increase in surface brightness as one moves slightly away from the core
is seen in this galaxy as well. The jets are noticeably asymmetric. The
relative distance of the flare with respect to the core is very different for
both the jets. The surface brightness of the northern jet decreases at a much
faster rate than that for the southern jet.  Another interesting feature is
the presence of wiggles along the tail. Although the jet is very asymmetric, it
is worth noting that every helix, also called wiggles, seen in one jet has a counterpart in the other jet.
These wiggles are very systematic and may be occurring because of the precession of
the central engine.

\begin{figure*}
\begin{center}
\begin{tabular}{c}
\includegraphics[width=18cm]{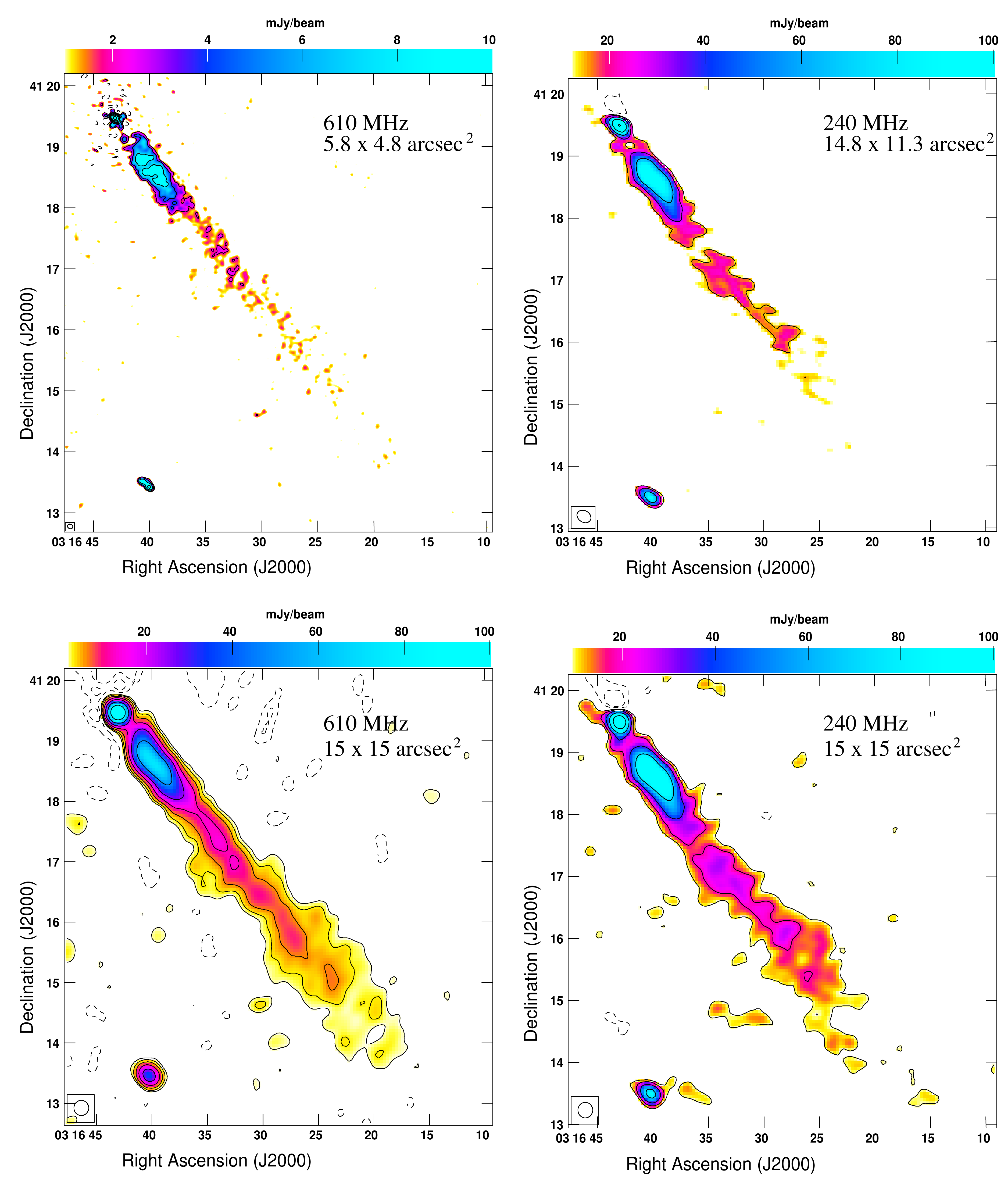}
\end{tabular}
\caption{Full synthesis images of IC\,310.
Top left panel: image at 610 MHz with a beam size of
%5.8$^{\prime\prime}$ $\times$ 4.8$^{\prime\prime}$ at a PA of 67.8$^\circ$,
$5\hbox{$.\!\!^{\prime\prime}$}8$ $\times$ $4\hbox{$.\!\!^{\prime\prime}$}8$ at a PA $67\hbox{$.\!\!^{\circ}$}8$,
the peak surface brightness is 168.0 mJy~beam$^{-1}$,
the error~bar at a source-free location is 0.3 mJy~beam$^{-1}$ and
the contour levels are 1.0 $\times$ ($-$2, 2, 4, 8, 16, 32, 64, 128)
mJy~beam$^{-1}$.
Top right panel: image at 240 MHz with a beam size of
%14.8$^{\prime\prime}$ $\times$ 11.3$^{\prime\prime}$ at a PA of $-$62.9$^\circ$,
$14\hbox{$.\!\!^{\prime\prime}$}8$ $\times$ $11\hbox{$.\!\!^{\prime\prime}$}3$ at a PA $-62\hbox{$.\!\!^{\circ}$}9$,
the peak surface brightness is 255.1 mJy~beam$^{-1}$,
the error~bar at a source-free location is 3.2 mJy~beam$^{-1}$ and
the contour levels are 7.6 $\times$ ($-$2, 2, 4, 8, 16, 32, 64)
mJy~beam$^{-1}$.
Bottom left panel: low-resolution image at 610 MHz with a beam size of
15$^{\prime\prime}$ $\times$ 15$^{\prime\prime}$, the peak surface brightness
is 180.5 mJy~beam$^{-1}$,
the error~bar at a source-free location is 0.8 mJy~beam$^{-1}$ and
the contour levels are 1.6 $\times$ ($-$2, $-$1, 1, 2, 4, 8, 16, 32, 64)
mJy~beam$^{-1}$.
Bottom right panel: low-resolution image at 240 MHz with a beam size of
15$^{\prime\prime}$ $\times$ 15$^{\prime\prime}$, the peak surface brightness
is 246.3 mJy~beam$^{-1}$, the error~bar at a source-free location is
3.4 mJy~beam$^{-1}$ and the contour levels are 9.0 $\times$ ($-$2, $-$1, 1, 2,
4, 8, 16, 32, 64) mJy~beam$^{-1}$.}
\label{ic310}
\end{center}
\end{figure*}

\begin{figure*}
\begin{center}
\begin{tabular}{c}
\includegraphics[width=18cm]{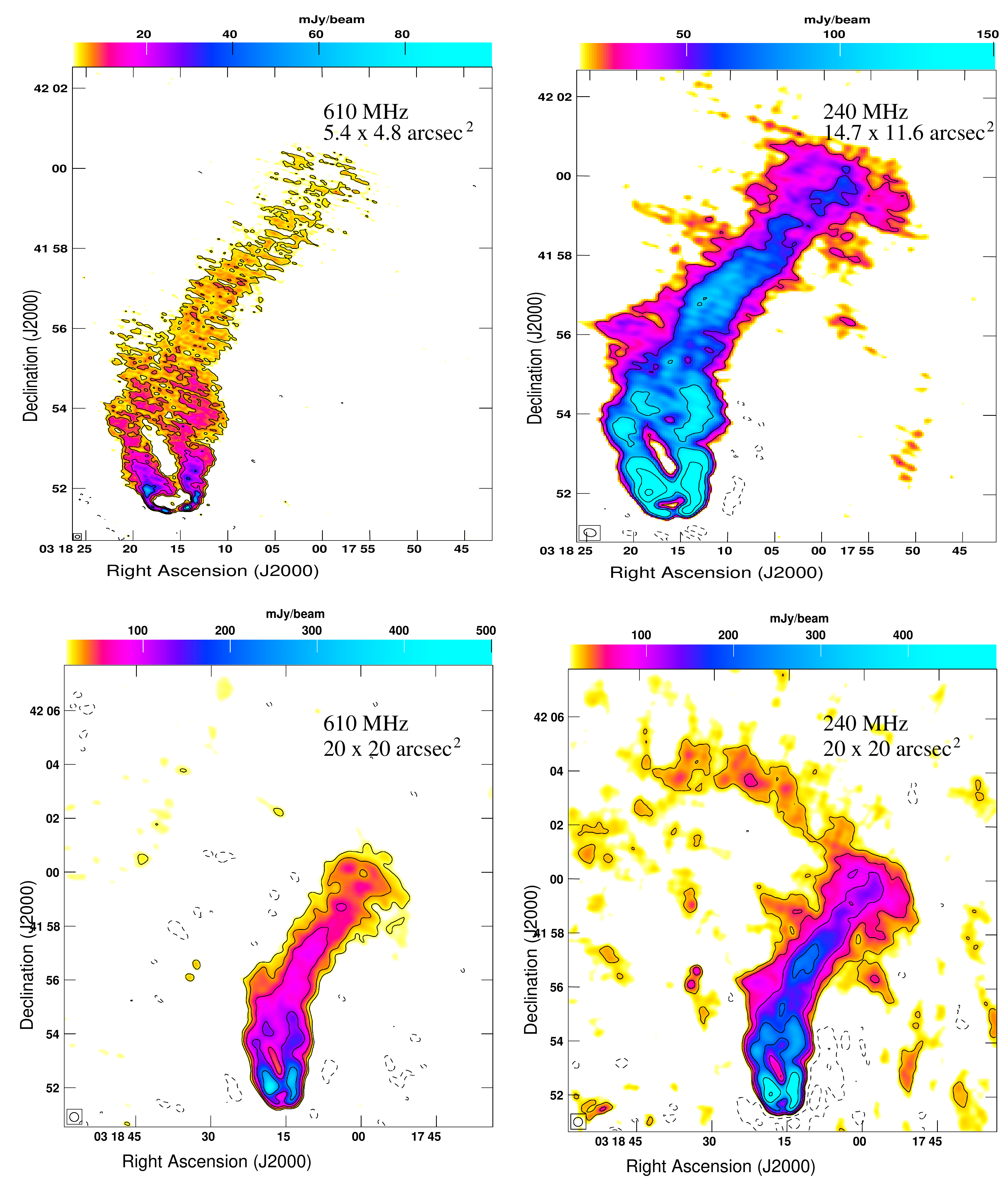}
\end{tabular}
\caption{Full synthesis images of NGC\,1265. Top left panel:
image at 610 MHz with a beam size of
%5.4$^{\prime\prime}$ $\times$ 4.8$^{\prime\prime}$ at a PA of $-$82.4$^\circ$,
$5\hbox{$.\!\!^{\prime\prime}$}4$ $\times$ $4\hbox{$.\!\!^{\prime\prime}$}8$ at a PA $-82\hbox{$.\!\!^{\circ}$}4$,
the peak surface brightness is 100.8 mJy~beam$^{-1}$,
the error~bar at a source-free location is 0.9 mJy~beam$^{-1}$ and
the contour levels are  1.0 $\times$ ($-$2, 2, 4, 8, 16, 32, 64, 128) mJy~beam$^{-1}$.
Top right panel:
image at 240 MHz with a beam size of
%14.7$^{\prime\prime}$ $\times$ 11.6$^{\prime\prime}$ at a PA of $-$68.9$^\circ$,
$14\hbox{$.\!\!^{\prime\prime}$}7$ $\times$ $11\hbox{$.\!\!^{\prime\prime}$}6$ at a PA $-68\hbox{$.\!\!^{\circ}$}9$,
the peak surface brightness is 338.3 mJy~beam$^{-1}$,
the error~bar at a source-free location is 4.4 mJy~beam$^{-1}$ and
the contour levels are  13.0 $\times$ ($-$2, 2, 4, 8, 16, 32, 64) mJy~beam$^{-1}$.
Bottom left panel:
low-resolution image at 610 MHz with
a beam size of 20$^{\prime\prime}$ $\times$ 20$^{\prime\prime}$,
the peak surface brightness is 299.2 mJy~beam$^{-1}$,
the error bar at a source-free location is 2.6 mJy~beam$^{-1}$ and
the contour levels are  15.0 $\times$ ($-$2, $-$1, 1, 2, 4, 8, 16, 32, 64, 128) mJy~beam$^{-1}$.
Bottom right panel:
low-resolution image at 240 MHz with
a beam size of 20$^{\prime\prime}$ $\times$ 20$^{\prime\prime}$,
the peak surface brightness is 467.0 mJy~beam$^{-1}$,
the error~bar at a source-free location is 6.4 mJy~beam$^{-1}$ and
the contour levels are  25.0 $\times$ ($-$2, $-$1, 1, 2, 4, 8, 16, 32, 64, 128) mJy~beam$^{-1}$.}
    \label{n1265}
\end{center}
\end{figure*}

\begin{figure*}
\begin{center}
\begin{tabular}{c}
\includegraphics[width=18cm]{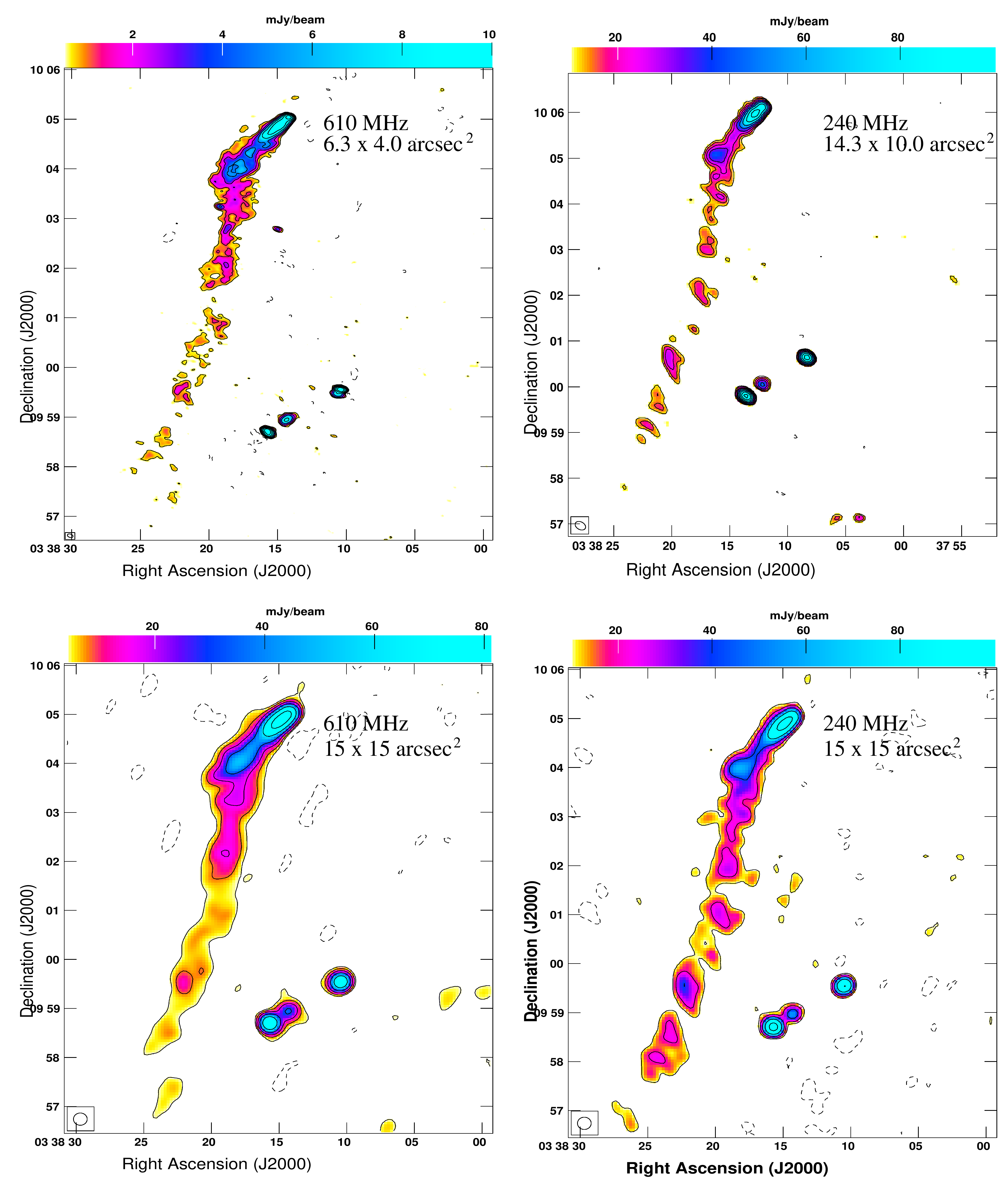}
\end{tabular}
\caption{Full synthesis images of GB6\,B0335$+$0955. Top left panel:
image at 610 MHz with
a beam size of
%6.3$^{\prime\prime}$ $\times$ 4.0$^{\prime\prime}$ at a PA of 69.8$^\circ$,
$6\hbox{$.\!\!^{\prime\prime}$}3$ $\times$ $4\hbox{$.\!\!^{\prime\prime}$}0$ at a PA $69\hbox{$.\!\!^{\circ}$}8$,
the peak surface brightness is 84.0 mJy~beam$^{-1}$,
the error~bar at a source-free location is 0.2 mJy~beam$^{-1}$ and
the contour levels are  0.6 $\times$ ($-$2, $-$1, 1, 2, 4, 8, 16, 32, 64) mJy~beam$^{-1}$.
Top right panel:
image at 240 MHz with
a beam size of
%14.3$^{\prime\prime}$ $\times$ 10.0$^{\prime\prime}$ at a PA of 59.9$^\circ$,
$14\hbox{$.\!\!^{\prime\prime}$}3$ $\times$ $10\hbox{$.\!\!^{\prime\prime}$}0$ at a PA $59\hbox{$.\!\!^{\circ}$}9$,
the peak surface brightness is 197.6 mJy~beam$^{-1}$,
the error~bar at a source-free location is 2.7 mJy~beam$^{-1}$ and
the contour levels are  11.0 $\times$ ($-$2, $-$1, 1, 2, 4, 8, 16).
Bottom left panel:
low-resolution image at 610 MHz with
a beam size of 15$^{\prime\prime}$ $\times$ 15$^{\prime\prime}$,
the peak surface brightness is 179.1 mJy~beam$^{-1}$,
the error~bar at a source-free location is 1.4 mJy~beam$^{-1}$ and
the contour levels are  4.0 $\times$ ($-$2, $-$1, 1, 2, 4, 8, 16, 32, 64) mJy~beam$^{-1}$.
Bottom right panel:
low-resolution image at 240 MHz with
a beam size of 15$^{\prime\prime}$ $\times$ 15$^{\prime\prime}$,
the peak surface brightness is 210.1 mJy~beam$^{-1}$,
the error~bar at a source-free location is 4.9 mJy~beam$^{-1}$ and
the contour levels are  10.0 $\times$ ($-$2, $-$1, 1, 2, 4, 8, 16, 32, 64) mJy~beam$^{-1}$.}
    \label{0335-0955}
\end{center}
\end{figure*}

\begin{figure*}
\begin{center}
\begin{tabular}{c}
\includegraphics[width=18cm]{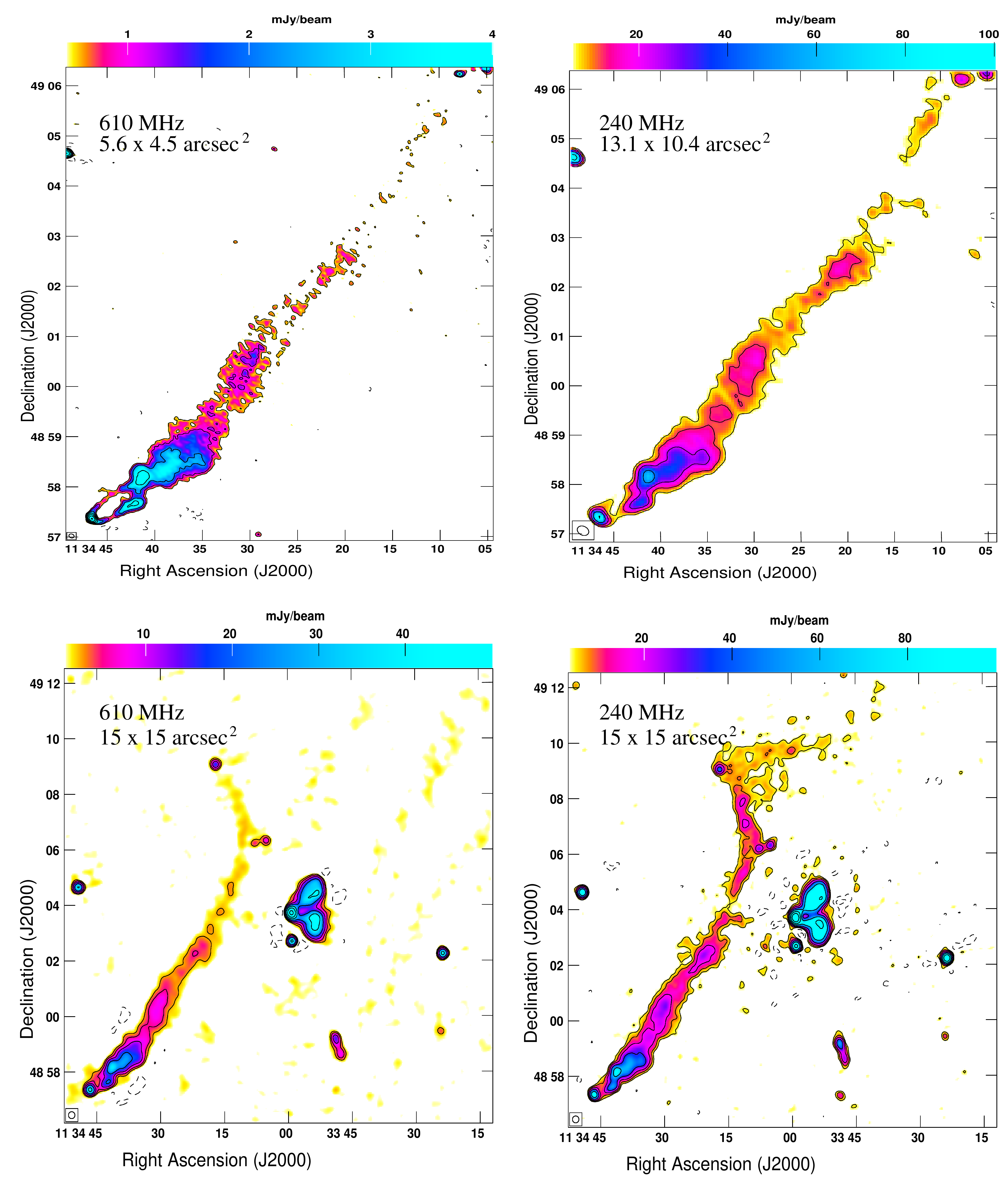}
\end{tabular}
\caption{Full synthesis images of IC\,711. Top left panel: image at 610
MHz with a beam size of
%5.6$^{\prime\prime}$ $\times$ 4.5$^{\prime\prime}$, at a PA of $-$61.4 $^\circ$,
$5\hbox{$.\!\!^{\prime\prime}$}6$ $\times$ $4\hbox{$.\!\!^{\prime\prime}$}5$ at a PA $-61\hbox{$.\!\!^{\circ}$}4$,
the peak surface brightness is 36.2 mJy~beam$^{-1}$,
the error bar found at a source-free location is 0.1 mJy~beam$^{-1}$ and the
contour levels are  0.3 $\times$ ($-$2, 2, 4, 8, 16, 32, 64, 128)
mJy~beam$^{-1}$. Top right panel: image at 240 MHz with a beam size of
%13.1$^{\prime\prime}$ $\times$ 10.4$^{\prime\prime}$, at a PA of 49.6 $^\circ$,
$13\hbox{$.\!\!^{\prime\prime}$}1$ $\times$ $10\hbox{$.\!\!^{\prime\prime}$}4$ at a PA $49\hbox{$.\!\!^{\circ}$}6$,
the peak surface brightness is 98.3 mJy~beam$^{-1}$, the error bar found at a
source-free location is 0.9 mJy~beam$^{-1}$ and the contour levels are  3.0
$\times$ ($-$2, 2, 4, 8, 16, 32, 64, 128, 256) mJy~beam$^{-1}$. Bottom
left panel:  low-resolution image at 610 MHz with a beam size of
15$^{\prime\prime}$ $\times$ 15$^{\prime\prime}$, the peak surface brightness
is 195.0 mJy~beam$^{-1}$, the error bar found at a source-free location is 0.6
mJy~beam$^{-1}$ and the contour levels are  2.5 $\times$ ($-$2, $-$1, 1, 2, 4,
8, 16, 32, 64) mJy~beam$^{-1}$. Bottom right panel:  low-resolution image
at 240 MHz with a beam size of 15$^{\prime\prime}$ $\times$
15$^{\prime\prime}$, the peak surface brightness is 508.0 mJy~beam$^{-1}$, the
error bar found at a source-free location is 0.9 mJy~beam$^{-1}$ and the
contour levels are  4.0 $\times$ ($-$2, $-$1, 1, 2, 4, 8, 16, 32, 64)
mJy~beam$^{-1}$.}
    \label{ic711}
\end{center}
\end{figure*}

\subsubsection{IC\,310}

The radio morphology of
IC\,310, located in the Perseus cluster, is shown in Figure~\ref{ic310}.
%For example, the classification of the IC\,310 as high-energy peaked BL~Lac

From an X-ray study of a few head$-$tail radio sources, \cite{1995MNRAS.277.1580E} had predicted the presence of a BL~Lac nucleus in IC\,310. Later on, \citet{2014A&A...563A..91A} classified IC\,310 as a high-energy peaked BL~Lac,
which suggests that the inclination angle of the radio jet
with respect to our line of sight is small. 
IC\,310 and NGC\,1265 in the Perseus cluster are two of the very well studied
head$-$tail radio galaxies
\citep{1968MNRAS.138....1R,1973A&A....26..413M,1986ApJ...301..841O, 2011ApJ...730...22P}. The head
is pointing toward the center of the cluster. The lower resolution images have
picked up more diffuse emission than have the higher resolution images,
as can be inferred from the contour maps (Figure~\ref{ic310}).
\cite{1998A&A...331..901S} and \cite{1998A&A...331..475F} have imaged the
sources in Perseus cluster at  1.4 GHz, 608 and 325 MHz. Their angular
resolution is much coarser than in our maps. However, they have recovered
more diffuse emission to the ends of the tails. Although they have mentioned
about evidence for a faint twin tail in high-frequency VLA maps, the resolution
of our maps does not seem to be enough to resolve them. The two tails are
so well aligned that the two tails can just not be distinguished from each
other. Since head$-$tail radio galaxies are seen in projection, it is hard to
conclude if they are indeed head$-$tail sources or only appear to be head$-$tail
sources in projection. However, there is a sudden increase in the surface
brightness and apparent spatial extent of the tail as one moves along the jet.
Recently IC\,310 was detected in $\gamma$-ray emission bands in the gigaelectronvolt and teraelectronvolt energy
ranges \citep{2010ATel.2510....1M}. \cite{2014A&A...563A..91A} ruled out
emission models occurring in a bow shock between the jet and the ICM, because
of the day-scale variability in the very high-energy band, and seemed to favor
the blazar-like scenario. Modeling of the spectral energy distribution of IC\,310
suggested the peaking of the synchrotron emission at X-ray wavelengths and
inverse-Compton radiation peaking in the multiteraelectronvolt band. Hence the source was
classified as a high-frequency peaked BL LAC. IC\,310 is the nearest known
blazar and the brightest radio galaxy at TeV energies.

\subsubsection{NGC\,1265}

The massive elliptical galaxy \citep[][Figure~\ref{all-ht}]{1986ApJ...301..841O} that marks the head of the
radio galaxy appears to be directed toward the center of the cluster. 

The high-resolution and low-resolution maps of NGC\,1265 at 610 and 240 MHz
are shown in Figure~\ref{n1265}. \cite{1986ApJ...301..841O} presented high
frequency images of NGC\,1265 and noted that the core possesses an inverted
spectrum, which explains why the core is not quite conspicuous in our
low-frequency maps while
it appears quite prominent in their high-frequency maps.  The individual jets
are clearly seen to be launched at wide angles in the high-resolution maps at
610 MHz. Farther downstream, the jets appear to be bent by a very large angle,
after which the jets expand in size and show higher surface brightness.
Subsequently, the radio jets develop wiggles, which were first reported by
\cite{1986ApJ...301..841O}.
% and are possibly caused due to precession or due to helical instabilities
%or due to variations in momentum flux.
\cite{1998A&A...331..901S} found extended emission to the northeastern side
with low-frequency observations and coarser resolutions. This extension was
found to have low surface brightness and a steep spectrum. Our map at 240 MHz,
which has a comparatively higher resolution, recovers a part of this extended
emission. 

\subsubsection{GB6\,B0335$+$0955}

Figure~\ref{all-ht} shows that the elliptical host galaxy of
GB6\,B0335$+$0955 is located at the tip of the
tail-like structure and is moving away from the cluster center. 

Figure~\ref{0335-0955} shows the radio morphologies of the head$-$tail
radio galaxy GB6\,B0335$+$0955 at 610 and 240 MHz.
The angular size of the galaxy
is about 8$^\prime$.5. The core is prominent, and the individual jets are not
resolved out. There is some evidence from the spatial profile of surface brightness near
the core for a dip in surface brightness and a brightening later on. 
%This galaxy is very similar to IC\,310.  
Farther down
the tail, one notices an abrupt change in the direction of motion of the galaxy
along with an increase in intensity. This could be because of interactions with neighboring galaxies that came close enough to cause deflection.
\cite{1995ApJ...451..125S} suggested that this was due to buoyant motions of
the tail following the low-density regions of the intergalactic medium.
The steepening of the spectrum as the distance increases from the host galaxy is evident
and is due to synchrotron aging of the radio-emitting plasma.

\subsubsection{IC\,711}
IC\,711 is the longest head$-$tail galaxy known
\citep{1987AJ.....94....1V,1988Ap&SS.149..225V,1988AJ.....95.1360V}. The high
resolution maps of IC\,711 at 610 and 240 MHz are presented in
Figure~\ref{ic711}.
Thin, unresolved jets are seen to emerge from the radio core.
Later on, both of the jets appear to widen into a high surface brightness
feature, similar to a hot spot. The low-resolution images reveal much more
extended features at the end of the tail. The galaxy seemed to have moved in an
orbital trajectory, about the centroid of the cluster whose position is marked
by the compact radio source IC\,712 to the east of the radio tail. There is a sudden
disruption toward the end of the tail. The low-resolution image at 240 MHz
shows much more emission toward the extreme ends of the tails, pointing to
comparatively high spectral index in that region. The morphology at the extreme
ends of the tails suggests a not-so-simple circular motion around the cluster
center \citep{2016arXiv161007783S}. However, it is also worth noting that the
point sources that mark various twists and jerks in the smooth structure are
not steep spectrum radio features, but instead are visible in both of the frequencies.  Farther
down the tail, the surface brightness drops, and the lower resolution image of
IC\,711 is similar to that of GB6\,B0335$+$0955 and IC\,310. 

\begin{figure*}
\begin{center}
\begin{tabular}{c}
\includegraphics[width=18cm]{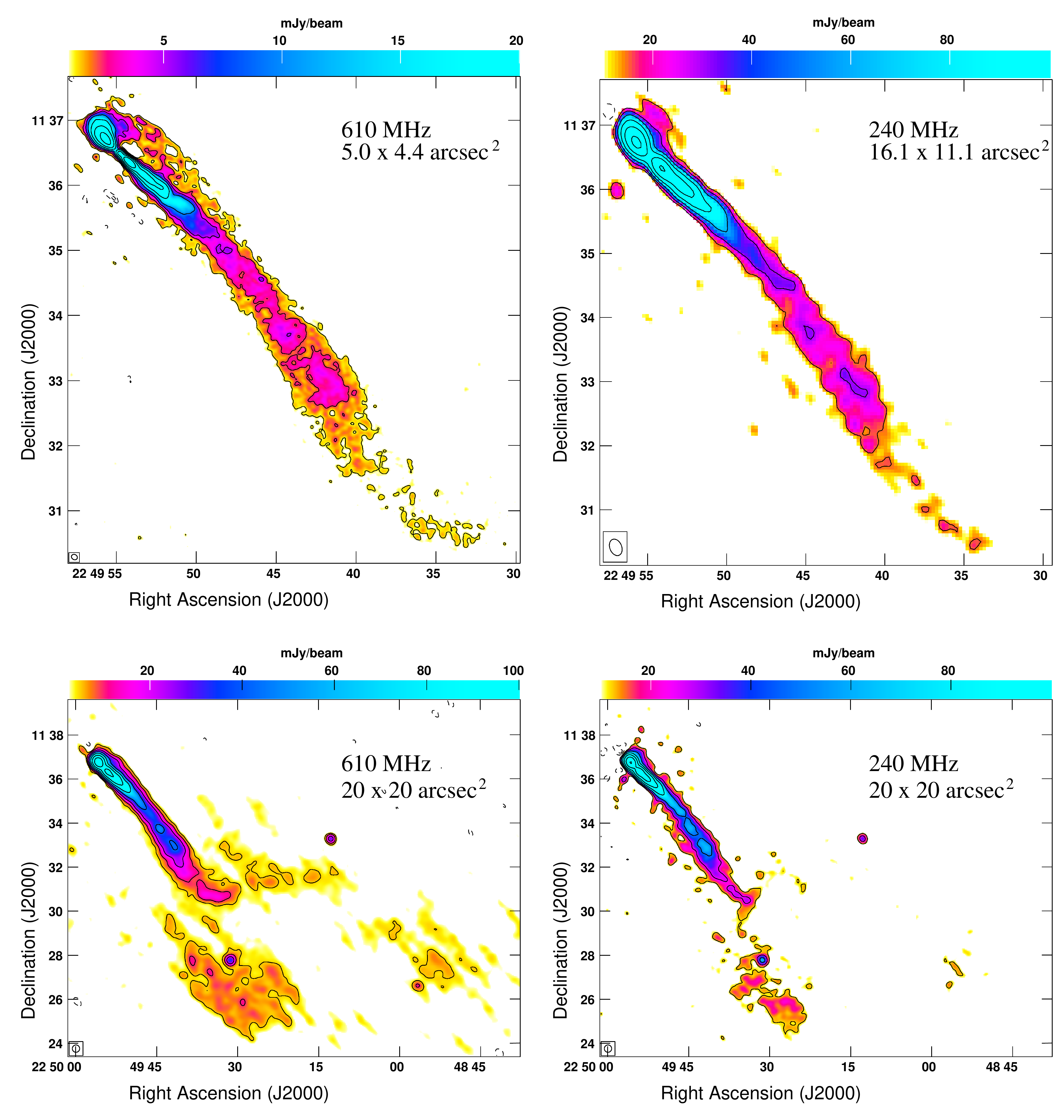}
\end{tabular}
\caption{Full synthesis images of NGC\,7385. Top left panel: image at 610
MHz with a beam size of
%5.0$^{\prime\prime}$ $\times$ 4.4$^{\prime\prime}$, at a PA of $-$60.2$^\circ$,
$5\hbox{$.\!\!^{\prime\prime}$}0$ $\times$ $4\hbox{$.\!\!^{\prime\prime}$}4$ at a PA $-60\hbox{$.\!\!^{\circ}$}2$,
the peak surface brightness is 146.2 mJy~beam$^{-1}$,
the error bar found at a source-free location is 0.2 mJy~beam$^{-1}$ and the
contour levels are  1.2 $\times$ ($-$2, $-$1, 1, 2, 4, 8, 16, 32, 64)
mJy~beam$^{-1}$. Top right panel: image at 240 MHz with a beam size of
%16.1$^{\prime\prime}$ $\times$ 11.08$^{\prime\prime}$, at a PA of $-$22.1$^\circ$,
$16\hbox{$.\!\!^{\prime\prime}$}1$ $\times$ $11\hbox{$.\!\!^{\prime\prime}$}1$ at a PA $-22\hbox{$.\!\!^{\circ}$}1$,
the peak surface brightness is 573.0 mJy~beam$^{-1}$, the
error bar found at a source-free location is 2.0 mJy~beam$^{-1}$ and the
contour levels are  15.0 $\times$ ($-$1, 1, 2, 4, 8, 16, 32, 64)
mJy~beam$^{-1}$. Bottom left panel:  low-resolution image at 610 MHz with
a beam size of 20$^{\prime\prime}$ $\times$ 20$^{\prime\prime}$, the peak
surface brightness is 461.3 mJy~beam$^{-1}$, the error bar found at a
source-free location is 1.4 mJy~beam$^{-1}$ and the contour levels are  6.0
$\times$ ($-$2, $-$1, 1, 2, 4, 8, 16, 32, 64) mJy~beam$^{-1}$. Bottom
right panel:  low-resolution image at 240 MHz with a beam size of
20$^{\prime\prime}$ $\times$ 20$^{\prime\prime}$, the peak surface brightness
is 667.8 mJy~beam$^{-1}$, the error bar found at a source-free location is 2.3
mJy~beam$^{-1}$ and the contour levels are  12.0 $\times$ ($-$2, $-$1, 1, 2, 4,
8, 16, 32, 64, 128) mJy~beam$^{-1}$.}
    \label{n7385}

\end{center}
\end{figure*}

\subsubsection{NGC\,7385}
\label{ng7385}

The location of the host galaxy of NGC\,7385 is shown in Figure~\ref{all-ht}.
The position of the optical host galaxy is not coincident with the tip of the
head$-$tail radio galaxy. This galaxy is a possible example where
the jet direction is not perpendicular to the direction of the
motion, as is seen in the classic examples of head$-$tail radio galaxies. However,
the peculiar morphology may also be explained using projection effects, as
was demonstrated by \cite{1980PASAu...4...74R}.

The radio maps of NGC\,7385 at 610 and 240 MHz are presented in
Figure~\ref{n7385}. \cite{1980ApJ...242..502H} had mapped this galaxy at 5 GHz
using the VLA.
The lower resolution images show
much more extended structures. Although the spectral index steepens along the
radio tail, the 240 MHz image at lower angular resolution does not show the
diffuse emission toward the ends of the tails. This is because of the poor
signal-to-noise ratio in the image, due to corruption from RFI. %It appears that
%the host galaxy has either changed its direction during its motion or undergone
%a change in the inclination angle; either way, this galaxy is not following a
%simple orbital motion around the cluster center.
NGC\,7385 is the only sample
head$-$tail radio galaxy in which an optical jet and also an X-ray jet were detected
\citep{2015MNRAS.452.3064R}.

\subsection{Spectral structure and physical parameters}

A population of relativistic electrons following a power law distribution of energies in the presence of a magnetic field
will emit synchrotron radiation, which follows a power-law in frequency. According to synchrotron theory,
the rate of energy loss of a radiating electron is proportional to the square of the energy of the electron.
Therefore, high-energy electrons lose energy faster and get depleted earlier than low-energy ones.
This depletion of high-energy electrons produces a break in the synchrotron spectrum, which moves to lower frequencies as time elapses. The spectrum also becomes steeper.
% also lead to a steepening of the spectrum.
In FR\,I radio galaxies, the electron population toward the end of the tail are much older than that near the radio core
since there is a constant injection of a fresh population of electrons near the core.
Additionally, the head$-$tail galaxies are moving through the ICM with a velocity comparable to the speed of the radio jet, as inferred from the bending \citep{1985ApJ...295...80O}.
The synchrotron-emitting plasma at the ends of the radio tails thus is stationary
with respect to the ICM.
Hence (1) it is harder for the newly injected electron population in head$-$tail radio galaxies to
reach the ends of the radio tails, and
(2) the spectral steepening in head$-$tail radio galaxies is expected to be more
pronounced than in normal FR\,I radio galaxies. 

\begin{figure}[t]
\begin{center}
\begin{tabular}{c}
\includegraphics[width=7.5cm]{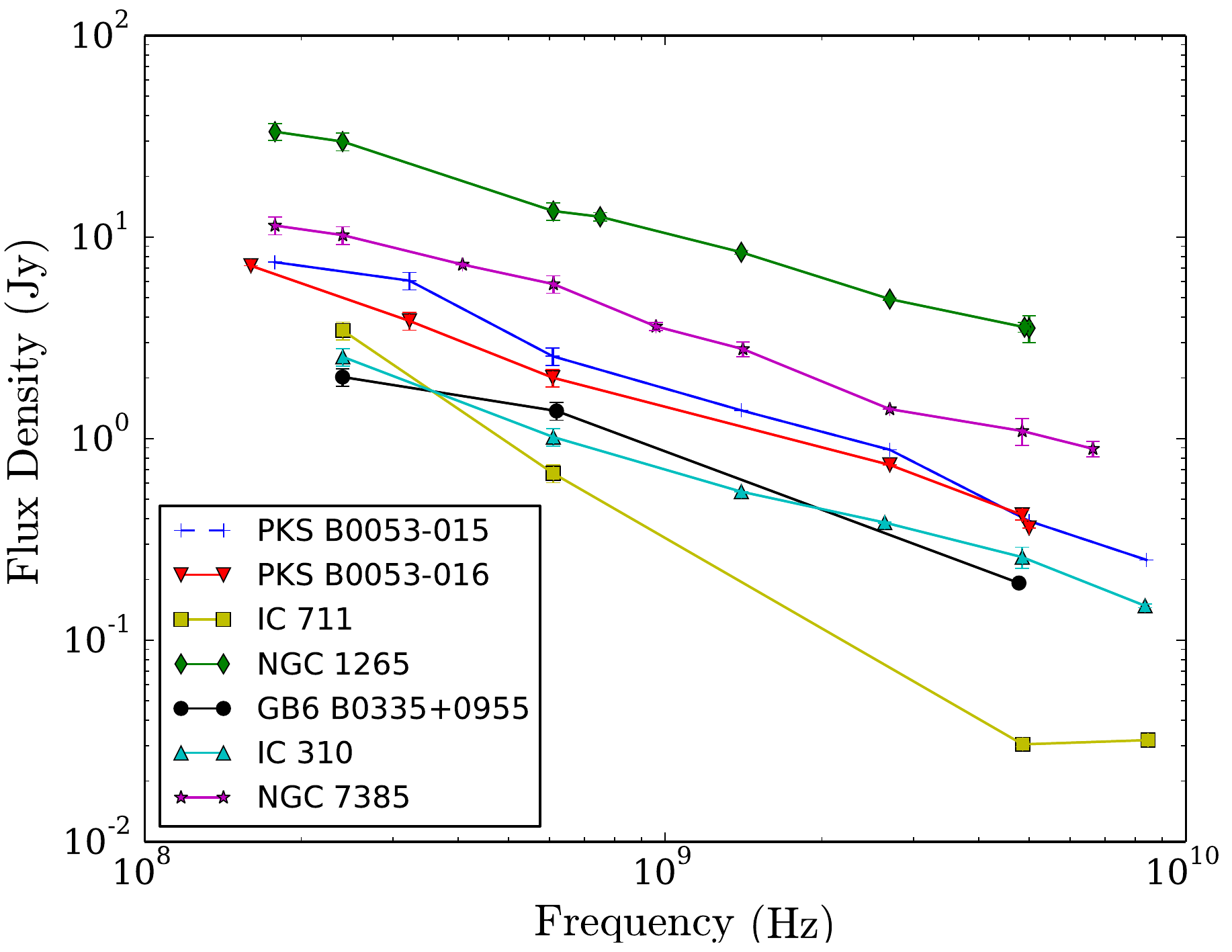}
\end{tabular}
\caption{Integrated flux densities all sources in the sample.
The GMRT flux density measurements at 240 MHz and 610 MHz
are in good agreement with the data at close frequencies taken from the
literature.
The error bars of the GMRT flux density measurements are quoted in
respective Figures (from Figure~\ref{0053-016} to Figures~\ref{n7385}).}
\label{int-flux}
\end{center}
\end{figure}

In order to calculate the rest-frame radio power, $L_{\textrm{610 MHz}}$,
we used the spectral index $\alpha$ between 240/325 and 610 MHz
(Figure~\ref{int-flux}) and
$$
L_{\textrm{610 MHz}}=4\pi D^2_L(z) \int S_{\nu} d \nu,
$$
where $D_L(z)$ is the luminosity distance and $S_{\textrm{610 MHz}}$ is the flux density at 610 MHz.

The role played by magnetic fields in the evolution and dynamics of head$-$tail
galaxies is barely explored.
Due to limited data, we estimate magnitudes of magnetic fields
using the minimum energy conditions \citep{1956ApJ...124..416B}.
The head$-$tail galaxies were divided into several regions of rectangular shape
perpendicular to the length of the tail, 
and the volumes of these regions were calculated assuming a cylindrical geometry.
Note that this could lead to underestimation, especially in cases where one of the tails lies over the other in projection.
We assume the electron population to radiate at frequencies from $10^{7}$ Hz to $10^{10}$ Hz
and we further assume that one-half of the total energy of the particles is carried by electrons.
We measure the total flux densities from both 240 and 610 MHz maps and determine
spectral indices for every region and estimate the equipartition magnetic field. A calibration error of 10$\%$ was assumed on flux density and was added to the RMS noise of the intensity maps in quadrature. 
Following \citet{Moffet1975}, we calculate the equipartition magnetic field $B_{\rm eq}$
using
$$
B_{\rm eq} = 2.3 (aAL/V)^{2/7};
$$
where $a$, $A$, $L$ and $V$ are the fraction of total energy in protons to that
in electrons, the shape factor, the total radio power, and the volume of the region,
respectively.  The radio spectra and the equipartition magnetic field as a
function of distance from the head are plotted in Figure~\ref{spec-in-eq-para}.
There is a clear decrease in the spectral index and in the magnetic field
values as one moves away from the core.
%The assumption that there is equipartition of energy between the radiating particles and magnetic field seems to break at certain regions in several sources, and the estimated radiative age peaks at these regions as compared to the adjacent regions.
%The spectral index is also plotted against the distance from the core.
Furthermore, the spectral index steepens when moving away from the center in all but two sources, IC\,310 and IC\,711
(Figure~\ref{spec-in-eq-para}).
%This trend is consistent with our expectations as was discussed in Section ~\ref{specstr}.
Briefly, the  electron population in the tails loses
a significant amount of its energy by radiation.
If the particles are in equipartition,
then the total energy will be shared almost equally as magnetic field energy
and electron energy.
Hence the equipartition magnetic field becomes weaker with distance from the head.
The profile of the minimum pressure,
$$
P_{\rm eq} = U_{\rm min}/3
$$
mimics the profile of the equipartition magnetic field of the radio plasma.
%The relatively flat spectra at two regions in the tails of IC\,310 and IC\,711
%(Figure~\ref{spec-in-eq-para}) suggest that
%the relativistic particles within them are continuously replenished.

%\subsubsection{Radiative ages}

We also derive the upper limit to the radiative age, assuming a magnetic field
%B~=~B$_{\rm IC}$/$\sqrt{3}$ and B$_{\rm IC}$ = 4~(1 + $z)^2$~$\mu$G,
${B_{\rm IC}}$ = 4~(1 + $z)^2$~$\mu$G,
of the synchrotron-emitting
electrons at these regions shown in Figure~\ref{spec-in-eq-para} using
$$
{t} = 1060 \frac{{B}^{0.5}}{({B}^2 +
{B}_{\rm IC}^2)}[(1 + z)\nu_{\rm br}]^{-0.5}~{\rm Myr}.
$$
Here, the frequency where the radio spectrum changes by 0.5 is
the break frequency, $\nu_{\rm br}$, in GHz, and
the magnetic fields, $B$ and $B_{\rm IC}$, are in $\mu$G \citep{Miley1980}.
Due to sparse radio frequency measurements of the
spatially resolved structures for our sample sources,
we assume $\nu_{\rm br} =$ 1420 MHz.
%The radiative age estimates are consistent with the hypothesis that the radio plasma ages due to synchrotron cooling as it moves away from the head of the radio galaxy and the radio plasma at the farthest end is the oldest.
%

\begin{figure*}
\begin{center}
\begin{tabular}{c}
\includegraphics[width=16.5cm]{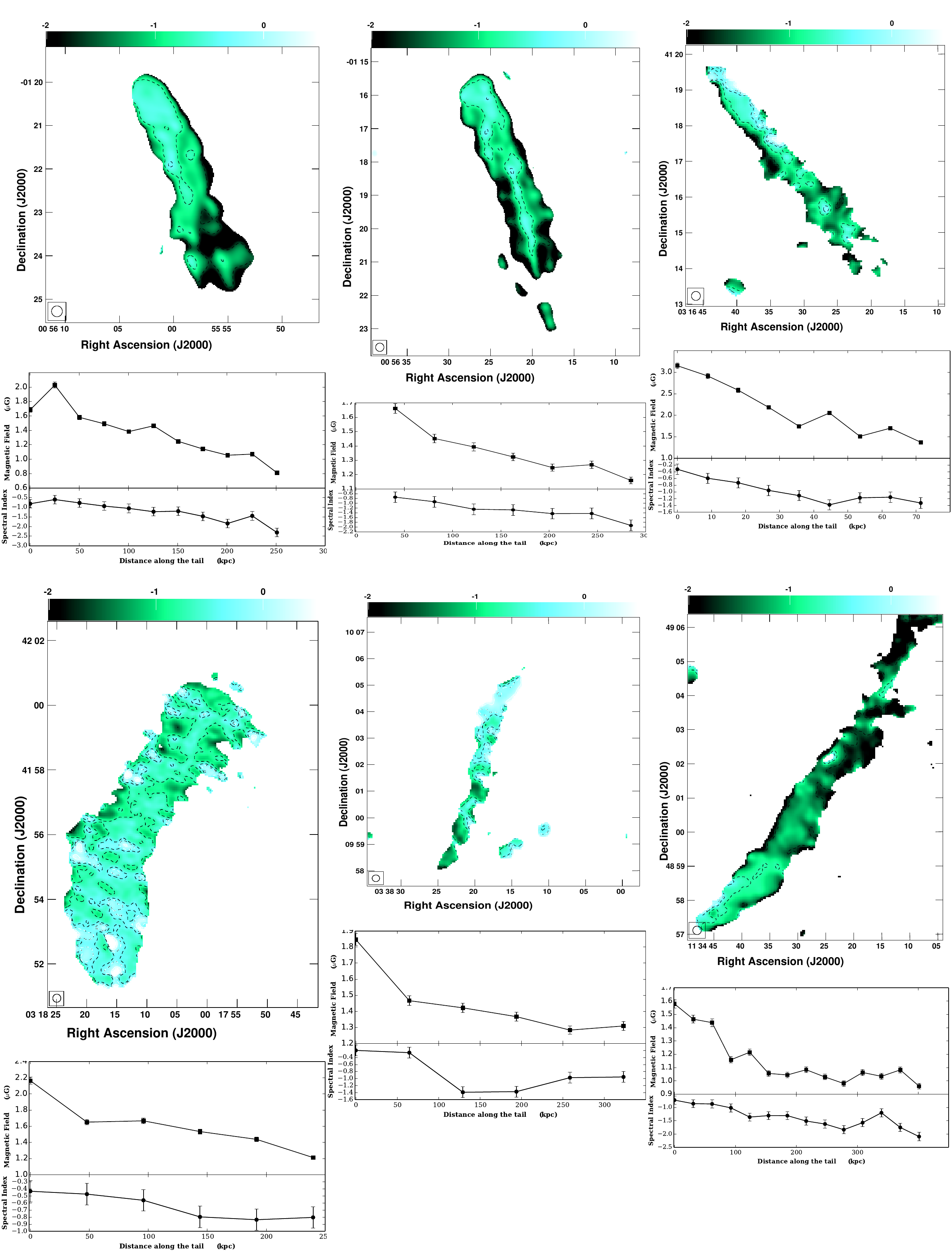}
\end{tabular} 
  \caption{Upper panel: distribution of spectral indices between 240/325 and 610 MHz
for PKS\,B0053$-$016 (left panel), PKS\,B0053$-$015 (middle panel) and
IC\,310 (right panel).
Lower panel: distribution of spectral indices, between 240 MHz and 610 MHz 
for NGC\,1265 (left panel), GB6\,B0335$+$0955 (middle panel) and
IC\,711 (right panel).
The corresponding panels below them are
profiles of the equipartition magnetic fields (upper panel) and
spectral index (lower panel) of the radio emitting plasma along the tail
from the head at the sampled positions of the radio galaxy.
In the spectral index maps, the lighter regions represent the
relatively flat spectrum regions as compared to
the darker regions which represent the steep spectrum.
The profile of the minimum pressure (not plotted; also see Table~\ref{phy-para}),
mimics the profile of the equipartition magnetic field of the radio plasma.}
  \label{spec-in-eq-para}
\end{center}
\end{figure*}

We present the measured source properties at the sampled locations
for each of our sample sources in
Table~\ref{phy-para} including
distance from the head along the tail (column~1), size (column~2), radio luminosity (column~3),
spectral index between 240 MHz (or 325 MHz) and 610 MHz (column~4),
equipartition parameters (from columns~5 to column~7) and radiative age (column~8).
Below we provide a description of the spectral structure
shown in Figure~\ref{spec-in-eq-para} from the radio
morphologies at 240/325 and 610 MHz discussed above for our sample sources.
Also presented in Figure~\ref{spec-in-eq-para} are profiles of the
equipartition parameters and spectral index of
the radio-emitting plasma along the tail from the head at the sample
positions of the radio galaxy.

\subsubsection{PKS\,B0053$-$016}

The spectral index value steepens from $-$0.82 in the region near the core to
$-$2.31 toward the farthest end of the tail. Apart from the steepening of the
spectral index, \cite{1999A&A...344..472F}
also finds evidence for a spectral curvature in the tails at about
a distance of 2$^{\prime}$ away from the core. Both the steepening and the
presence of spectral curvature are consistent with the aging of the plasma
toward the ends of the tails.
Figure~\ref{spec-in-eq-para} shows that
there is an increase in the spectral index
next to the core. Although this discontinuity in the steepening trend of the
spectral index just after the core is within the error bars, there are certain
factors worth noticing. The surface brightness of the radio plasma in this
region suddenly increases and then decreases downstream, as can be inferred from
Table~\ref{phy-para}. The flatter spectrum compared to that near the core
indicates that some of the electrons were pushed to higher energies.  The net
increase in energy of the electrons leads to an increase in energy stored in the
magnetic field if minimum energy conditions hold true. This leads to the
noticeable change in the trend in the magnetic field. There is a weak
discontinuity in the magnetic field farther downstream even though there are no
such signatures in the spectral index variation. This is probably due to an
underestimation of the volume and may not be a real physical trend.

\subsubsection{PKS\,B0053$-$015}

Although the spectral index and magnetic field steepen without any noticeable
discontinuities, the spectral index map shows several interesting features. In
almost all the head$-$tail galaxies presented here, there is a steepening trend in the
spectral index as one traverses away from the head along the tail.
%As mentioned in the Section.~\ref{pksbmorph}, the projected jet direction is not
%perpendicular to the length of the radio galaxy or the direction of motion of the galaxy.

 But, in PKS\,0053$-$015, there is a region where a narrow filament of less-steep plasma seems to be embedded in a steeper background. The surface
brightness maps (see Figure~\ref{0056-015}) show that the narrow flat spectrum
filament coincides with the jet that is directed toward the south, while the
diffuser background is more likely to correspond to the jet that is directed
northward. They may not be spatially coinciding but appear so in projection.
This difference in the spectral index points to differently aged electron
populations. The fact that the steep background that corresponds to the
northern jet is more extended also points toward a more dynamical evolution.
Also, in Section.~\ref{pksbmorph} it was pointed out that the jets are launched
at an angle smaller than $90^\circ$ with the direction of motion of the galaxy.
Hence, the difference in the ages of the plasma indicates that this peculiar
orientation of jets must have remained the same throughout the entire duration
when the host galaxy traversed the length of the radio emission, which
represents its past trajectory. 
%\cite{1999A&A...344..472F} estimates 
\setcounter{figure}{9}
\begin{figure}
\begin{center}
\begin{tabular}{c}
\includegraphics[width=6.0cm]{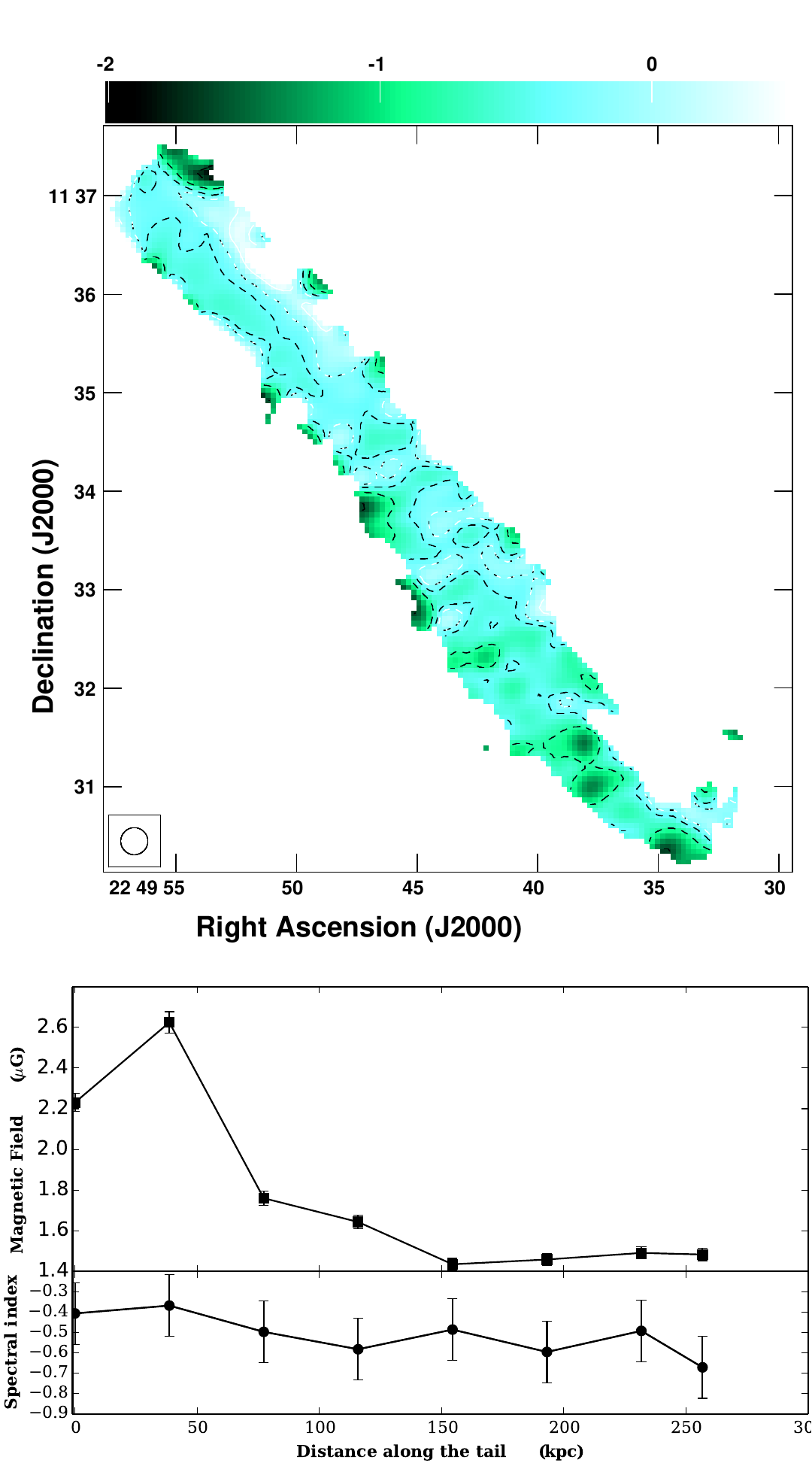}
\end{tabular} 
  \caption{\textit{Continued} -- The distribution of spectral indices, between 240 MHz and 610 MHz
for NGC\,7385 .}%-- \textit{Continued}
  \label{spec-in-eq-para-b}
\end{center}
\end{figure}

\begin{table*}
  \centering
  \caption{Distance from the head along the tail, size, radio luminosity, and
equipartition parameters along with radiative age at the sampled
position for the sample source calculated following the formalism explained in Sect.~\ref{ht-discuss}.}
  \label{phy-para}
\begin{tabular}{crrccrrr}
\hline
\hline
Distance &  \multicolumn{1}{c}{Size} & \multicolumn{1}{c}{Luminosity} & $\alpha$ & $B_{\rm min}$ & \multicolumn{1}{c}{$U_{\rm min}$} & \multicolumn{1}{c}{$P_{\rm min}$} & \multicolumn{1}{c}{Age} \\
(kpc) &  \multicolumn{1}{c}{(cm$^{-3}$)}  & \multicolumn{1}{c}{(erg~s$^{-1}$)} &  & ($\mu$G)  &   \multicolumn{1}{c}{(erg~cm$^{-3}$)} & \multicolumn{1}{c}{(dyne~cm$^{-2}$)} & \multicolumn{1}{c}{(Myr)} \\
 (1) &  \multicolumn{1}{c}{(2)} &  \multicolumn{1}{c}{(3)} & (4) & (5) &  \multicolumn{1}{c}{(6)} & \multicolumn{1}{c}{(7)} & \multicolumn{1}{c}{(8)} \\
\hline

\multicolumn{8}{l}{PKS\,0053$-$016} \\
~~0.0 & 6.5$\times10^{68}$ & 2.3 $\pm$0.3 $\times$ $10^{39}$ & $-$0.82 & 1.7 &11.3 $\pm$0.5 $\times$ $10^{-14}$ & 3.8 $\times$ $10^{-14}$ & 52.2 \\
~25.1 &12.3$\times10^{68}$ & 8.0 $\pm$1.2 $\times$ $10^{39}$ & $-$0.60 & 2.0 &16.4 $\pm$0.7 $\times$ $10^{-14}$ & 5.5 $\times$ $10^{-14}$ & 53.9 \\
~50.2 &16.6$\times10^{68}$ & 4.6 $\pm$0.8 $\times$ $10^{39}$ & $-$0.77 & 1.6 & 9.9 $\pm$0.4 $\times$ $10^{-14}$ & 3.3 $\times$ $10^{-14}$ & 51.5 \\
~75.3 &15.6$\times10^{68}$ & 3.4 $\pm$0.5 $\times$ $10^{39}$ & $-$0.94 & 1.5 & 8.9 $\pm$0.4 $\times$ $10^{-14}$ & 3.0 $\times$ $10^{-14}$ & 50.6 \\
100.4 &16.9$\times10^{68}$ & 2.9 $\pm$0.5 $\times$ $10^{39}$ & $-$1.06 & 1.4 & 7.6 $\pm$0.3 $\times$ $10^{-14}$ & 2.5 $\times$ $10^{-14}$ & 49.5 \\
125.4 &13.2$\times10^{68}$ & 2.9 $\pm$0.6 $\times$ $10^{39}$ & $-$1.23 & 1.5 & 8.5 $\pm$0.4 $\times$ $10^{-14}$ & 2.8 $\times$ $10^{-14}$ & 50.6 \\
150.5 &19.3$\times10^{68}$ & 2.4 $\pm$0.4 $\times$ $10^{39}$ & $-$1.20 & 1.3 & 6.2 $\pm$0.3 $\times$ $10^{-14}$ & 2.1 $\times$ $10^{-14}$ & 48.3 \\
175.6 &21.6$\times10^{68}$ & 1.9 $\pm$0.3 $\times$ $10^{39}$ & $-$1.47 & 1.1 & 5.2 $\pm$0.2 $\times$ $10^{-14}$ & 1.7 $\times$ $10^{-14}$ & 45.5 \\
200.7 &28.6$\times10^{68}$ & 1.8 $\pm$0.4 $\times$ $10^{39}$ & $-$1.84 & 1.1 & 4.4 $\pm$0.2 $\times$ $10^{-14}$ & 1.5 $\times$ $10^{-14}$ & 45.5 \\
225.8 &32.2$\times10^{68}$ & 2.2 $\pm$0.4 $\times$ $10^{39}$ & $-$1.45 & 1.1 & 4.6 $\pm$0.2 $\times$ $10^{-14}$ & 1.5 $\times$ $10^{-14}$ & 45.5 \\
250.9 &32.9$\times10^{68}$ & 9.4 $\pm$1.6 $\times$ $10^{38}$ & $-$2.31 & 0.8 & 2.6 $\pm$0.1 $\times$ $10^{-14}$ & 0.9 $\times$ $10^{-14}$ & 40.0 \\
\hline                                                                                                                                           
                                                                                                                                                 
\multicolumn{8}{l}{PKS\,0053$-$015} \\                                                                                                           
~~0.0 & 2.3$\times10^{69}$ & 7.6 $\pm$1.1 $\times$ $10^{39}$ & $-$0.75 & 1.7 & 1.10 $\pm$0.5 $\times$ $10^{-14}$ & 0.3 $\times$ $10^{-14}$& 53.0  \\
~40.7 & 3.7$\times10^{69}$ & 7.5 $\pm$1.2 $\times$ $10^{39}$ & $-$0.94 & 1.5 &  8.4 $\pm$0.3 $\times$ $10^{-14}$ & 2.8 $\times$ $10^{-14}$& 51.3 \\
~81.3 & 3.2$\times10^{69}$ & 5.9 $\pm$1.1 $\times$ $10^{39}$ & $-$1.25 & 1.4 &  7.7 $\pm$0.3 $\times$ $10^{-14}$ & 2.6 $\times$ $10^{-14}$& 50.3  \\
122.0 & 2.9$\times10^{69}$ & 4.3 $\pm$0.8 $\times$ $10^{39}$ & $-$1.28 & 1.3 &  7.0 $\pm$0.3 $\times$ $10^{-14}$ & 2.3 $\times$ $10^{-14}$& 49.1 \\
162.7 & 3.1$\times10^{69}$ & 3.8 $\pm$0.7 $\times$ $10^{39}$ & $-$1.44 & 1.3 &  6.2 $\pm$0.3 $\times$ $10^{-14}$ & 2.1 $\times$ $10^{-14}$& 49.1  \\
203.4 & 3.0$\times10^{69}$ & 3.8 $\pm$0.8 $\times$ $10^{39}$ & $-$1.43 & 1.3 &  6.4 $\pm$0.3 $\times$ $10^{-14}$ & 2.1 $\times$ $10^{-14}$& 49.1 \\
244.0 & 3.0$\times10^{69}$ & 2.8 $\pm$0.5 $\times$ $10^{39}$ & $-$1.93 & 1.2 &  5.3 $\pm$0.2 $\times$ $10^{-14}$ & 1.8 $\times$ $10^{-14}$& 47.7  \\
\hline                                                                                                                                           
                                                                                                                                                 
\multicolumn{8}{l}{IC\,310} \\                                                                                                        
~~0.0 & 1.2$\times10^{68}$ & 6.4$\pm$2.0 $\times$ $10^{38}$ & $-$0.33 & 1.9 & 15.1 $\pm$0.6 $\times$ $10^{-14}$ & 5.0 $\times$ $10^{-14}$ &58.2\\
~17.4 & 1.8$\times10^{68}$ &10.6$\pm$2.7 $\times$ $10^{38}$ & $-$0.59 & 2.0 & 16.4 $\pm$0.6 $\times$ $10^{-14}$ & 5.5 $\times$ $10^{-14}$ &58.7\\
~36.4 & 1.8$\times10^{68}$ & 8.1$\pm$2.2 $\times$ $10^{38}$ & $-$0.73 & 1.8 & 13.3 $\pm$0.5 $\times$ $10^{-14}$ & 4.4 $\times$ $10^{-14}$ &57.7\\
~55.5 & 2.0$\times10^{68}$ & 3.6$\pm$1.2 $\times$ $10^{38}$ & $-$0.95 & 1.4 &  8.1 $\pm$0.3 $\times$ $10^{-14}$ & 2.7 $\times$ $10^{-14}$ &54.3\\
~74.5 & 3.3$\times10^{68}$ & 3.7$\pm$0.8 $\times$ $10^{38}$ & $-$1.11 & 1.2 &  6.2 $\pm$0.2 $\times$ $10^{-14}$ & 2.1 $\times$ $10^{-14}$ &51.7\\
~93.5 & 2.2$\times10^{68}$ & 3.5$\pm$0.6 $\times$ $10^{38}$ & $-$1.38 & 1.3 &  7.2 $\pm$0.2 $\times$ $10^{-14}$ & 2.4 $\times$ $10^{-14}$ &53.1\\
112.0 & 5.0$\times10^{68}$ & 3.6$\pm$0.6 $\times$ $10^{38}$ & $-$1.17 & 1.1 &  5.0 $\pm$0.2 $\times$ $10^{-14}$ & 1.7 $\times$ $10^{-14}$ &50.1\\
131.5 & 3.8$\times10^{68}$ & 2.9$\pm$0.5 $\times$ $10^{38}$ & $-$1.16 & 1.1 &  4.9 $\pm$0.2 $\times$ $10^{-14}$ & 1.6 $\times$ $10^{-14}$ &50.1\\
150.0 & 3.6$\times10^{68}$ & 1.6$\pm$0.3 $\times$ $10^{38}$ & $-$1.33 & 0.9 &  3.9 $\pm$0.1 $\times$ $10^{-14}$ & 1.3 $\times$ $10^{-14}$ &46.3\\
\hline
%
% ~0.0 & 2.0$\times10^{67}$ & 6.4 $\pm$2.0 $\times$ $10^{38}$ & $-$0.33 & 3.2 &39.5 $\pm$1.6 $\times$ $10^{-14}$ &13.2 $\times$ $10^{-14}$ & 57.4 \\
% ~8.9 & 5.0$\times10^{67}$ &10.6 $\pm$2.7 $\times$ $10^{38}$ & $-$0.59 & 2.9 &33.8 $\pm$1.4 $\times$ $10^{-14}$ &11.3 $\times$ $10^{-14}$ & 58.5 \\
% 17.8 & 5.0$\times10^{67}$ & 8.1 $\pm$2.2 $\times$ $10^{38}$ & $-$0.73 & 2.6 &26.5 $\pm$1.1 $\times$ $10^{-14}$ & 8.9 $\times$ $10^{-14}$ & 59.2 \\
% 26.8 & 4.0$\times10^{67}$ & 3.6 $\pm$1.2 $\times$ $10^{38}$ & $-$0.95 & 2.2 &19.0 $\pm$0.8 $\times$ $10^{-14}$ & 6.3 $\times$ $10^{-14}$ & 59.2 \\
% 35.7 &10.0$\times10^{67}$ & 3.7 $\pm$0.8 $\times$ $10^{38}$ & $-$1.11 & 1.8 &12.1 $\pm$0.5 $\times$ $10^{-14}$ & 4.0 $\times$ $10^{-14}$ & 57.7 \\
% 44.6 & 5.0$\times10^{67}$ & 3.5 $\pm$0.6 $\times$ $10^{38}$ & $-$1.38 & 2.1 &16.8 $\pm$0.7 $\times$ $10^{-14}$ & 5.6 $\times$ $10^{-14}$ & 59.0 \\
% 53.5 &17.0$\times10^{67}$ & 3.6 $\pm$0.6 $\times$ $10^{38}$ & $-$1.17 & 1.5 & 9.1 $\pm$0.4 $\times$ $10^{-14}$ & 3.0 $\times$ $10^{-14}$ & 55.4 \\
% 62.5 & 9.0$\times10^{67}$ & 2.9 $\pm$0.5 $\times$ $10^{38}$ & $-$1.16 & 1.7 &11.5 $\pm$0.5 $\times$ $10^{-14}$ & 3.8 $\times$ $10^{-14}$ & 57.1 \\
% 71.4 &12.0$\times10^{67}$ & 1.6 $\pm$0.3 $\times$ $10^{38}$ & $-$1.33 & 1.4 & 7.5 $\pm$0.3 $\times$ $10^{-14}$ & 2.5 $\times$ $10^{-14}$ & 54.3 \\
%\hline                                                                                                                                           
                                                                                                                                                 
\multicolumn{8}{l}{NGC\,1265} \\                                                                                                                 
~~0.0 & 2.9$\times10^{69}$ & 2.2 $\pm$0.5 $\times$ $10^{40}$ & $-$0.43 & 2.2 &18.6 $\pm$0.8 $\times$ $10^{-14}$ & 6.2 $\times$ $10^{-14}$ & 57.9 \\
~47.9 & 7.0$\times10^{69}$ & 2.1 $\pm$0.4 $\times$ $10^{40}$ & $-$0.47 & 1.7 &10.8 $\pm$0.4 $\times$ $10^{-14}$ & 3.6 $\times$ $10^{-14}$ & 55.7 \\
~95.9 & 5.0$\times10^{69}$ & 1.5 $\pm$0.3 $\times$ $10^{40}$ & $-$0.56 & 1.7 &11.1 $\pm$0.5 $\times$ $10^{-14}$ & 3.7 $\times$ $10^{-14}$ & 55.7 \\
143.8 & 4.5$\times10^{69}$ & 1.0 $\pm$0.2 $\times$ $10^{40}$ & $-$0.80 & 1.5 & 9.4 $\pm$0.4 $\times$ $10^{-14}$ & 3.1 $\times$ $10^{-14}$ & 54.0 \\
191.8 & 4.9$\times10^{69}$ & 9.0 $\pm$1.5 $\times$ $10^{39}$ & $-$0.84 & 1.4 & 8.2 $\pm$0.3 $\times$ $10^{-14}$ & 2.7 $\times$ $10^{-14}$ & 53.0 \\
239.7 & 8.2$\times10^{69}$ & 8.3 $\pm$1.5 $\times$ $10^{39}$ & $-$0.80 & 1.2 & 5.8 $\pm$0.2 $\times$ $10^{-14}$ & 1.9 $\times$ $10^{-14}$ & 50.4 \\
\hline                                                                                                                                           
                                                                                                                                                 
\multicolumn{8}{l}{GB6\,B0335$+$0955} \\                                                                                                         
~~0.0 & 1.5$\times10^{69}$ & 7.1 $\pm$1.4 $\times$ $10^{39}$ & $-$0.20 & 1.9 &13.6 $\pm$0.6 $\times$ $10^{-14}$ & 4.5 $\times$ $10^{-14}$ & 54.2 \\
~64.6 & 2.5$\times10^{69}$ & 4.8 $\pm$1.0 $\times$ $10^{39}$ & $-$0.26 & 1.5 & 8.6 $\pm$0.4 $\times$ $10^{-14}$ & 2.9 $\times$ $10^{-14}$ & 51.3 \\
129.3 & 1.5$\times10^{69}$ & 3.7 $\pm$0.6 $\times$ $10^{39}$ & $-$1.39 & 1.4 & 8.0 $\pm$0.3 $\times$ $10^{-14}$ & 2.7 $\times$ $10^{-14}$ & 50.3 \\
193.9 & 1.3$\times10^{69}$ & 2.2 $\pm$0.4 $\times$ $10^{39}$ & $-$1.37 & 1.4 & 7.4 $\pm$0.3 $\times$ $10^{-14}$ & 2.5 $\times$ $10^{-14}$ & 50.3 \\
258.6 & 1.5$\times10^{69}$ & 1.8 $\pm$0.4 $\times$ $10^{39}$ & $-$0.97 & 1.3 & 6.6 $\pm$0.3 $\times$ $10^{-14}$ & 2.2 $\times$ $10^{-14}$ & 49.1 \\
323.2 & 1.1$\times10^{69}$ & 1.3 $\pm$0.2 $\times$ $10^{39}$ & $-$0.95 & 1.3 & 6.8 $\pm$0.3 $\times$ $10^{-14}$ & 2.3 $\times$ $10^{-14}$ & 49.1 \\
\hline                                                                                                                                           
                                                                                                                                                 
\multicolumn{8}{l}{IC\,711} \\                                                                                                                   
~~0.0 & 3.3$\times10^{68}$ & 7.9 $\pm$1.8 $\times$ $10^{38}$ & $-$0.72 & 1.6 & 9.9 $\pm$0.4 $\times$ $10^{-14}$ & 3.3 $\times$ $10^{-14}$ & 53.4 \\
~30.9 & 7.3$\times10^{68}$ &13.9 $\pm$2.6 $\times$ $10^{38}$ & $-$0.86 & 1.5 & 8.5 $\pm$0.4 $\times$ $10^{-14}$ & 2.8 $\times$ $10^{-14}$ & 52.5 \\
~61.7 &13.7$\times10^{68}$ &24.0 $\pm$5.0 $\times$ $10^{38}$ & $-$0.87 & 1.4 & 8.2 $\pm$0.3 $\times$ $10^{-14}$ & 2.7 $\times$ $10^{-14}$ & 51.4 \\
~92.6 &18.8$\times10^{68}$ &15.8 $\pm$3.3 $\times$ $10^{38}$ & $-$1.01 & 1.2 & 5.3 $\pm$0.2 $\times$ $10^{-14}$ & 1.8 $\times$ $10^{-14}$ & 48.9 \\
123.5 & 7.9$\times10^{68}$ & 7.8 $\pm$1.6 $\times$ $10^{38}$ & $-$1.36 & 1.2 & 5.9 $\pm$0.2 $\times$ $10^{-14}$ & 2.0 $\times$ $10^{-14}$ & 48.9 \\
154.4 &18.1$\times10^{68}$ &10.9 $\pm$2.0 $\times$ $10^{38}$ & $-$1.30 & 1.1 & 4.4 $\pm$0.2 $\times$ $10^{-14}$ & 1.5 $\times$ $10^{-14}$ & 47.4 \\
185.2 &13.7$\times10^{68}$ & 8.0 $\pm$1.4 $\times$ $10^{38}$ & $-$1.31 & 1.0 & 4.3 $\pm$0.2 $\times$ $10^{-14}$ & 1.4 $\times$ $10^{-14}$ & 45.6 \\
216.1 & 7.0$\times10^{68}$ & 4.7 $\pm$0.8 $\times$ $10^{38}$ & $-$1.51 & 1.1 & 4.7 $\pm$0.2 $\times$ $10^{-14}$ & 1.6 $\times$ $10^{-14}$ & 47.4 \\
245.0 &11.7$\times10^{68}$ & 6.2 $\pm$1.3 $\times$ $10^{38}$ & $-$1.62 & 1.0 & 4.2 $\pm$0.2 $\times$ $10^{-14}$ & 1.4 $\times$ $10^{-14}$ & 45.6 \\
277.9 &12.1$\times10^{68}$ & 5.7 $\pm$1.1 $\times$ $10^{38}$ & $-$1.83 & 1.0 & 3.8 $\pm$0.2 $\times$ $10^{-14}$ & 1.3 $\times$ $10^{-14}$ & 45.6 \\
308.7 & 5.1$\times10^{68}$ & 3.0 $\pm$0.6 $\times$ $10^{38}$ & $-$1.57 & 1.1 & 4.5 $\pm$0.2 $\times$ $10^{-14}$ & 1.5 $\times$ $10^{-14}$ & 47.4 \\
339.6 & 4.4$\times10^{68}$ & 2.3 $\pm$0.8 $\times$ $10^{38}$ & $-$1.20 & 1.0 & 4.3 $\pm$0.2 $\times$ $10^{-14}$ & 1.4 $\times$ $10^{-14}$ & 45.6 \\
370.5 & 4.3$\times10^{68}$ & 2.9 $\pm$0.7 $\times$ $10^{38}$ & $-$1.75 & 1.1 & 4.7 $\pm$0.2 $\times$ $10^{-14}$ & 1.6 $\times$ $10^{-14}$ & 47.4 \\
401.4 & 5.4$\times10^{68}$ & 2.2 $\pm$0.5 $\times$ $10^{38}$ & $-$2.09 & 1.0 & 3.7 $\pm$0.2 $\times$ $10^{-14}$ & 1.2 $\times$ $10^{-14}$ & 45.6 \\
\hline                                                                                                                                           
                                                                                                                                                 
\multicolumn{8}{l}{NGC\,7385} \\                                                                                                                 
~~0.0 &11.3$\times10^{68}$ & 9.7 $\pm$2.3 $\times$ $10^{39}$ & $-$0.41 & 2.2 &19.8 $\pm$0.8 $\times$ $10^{-14}$ & 6.6 $\times$ $10^{-14}$ & 57.7 \\
~38.6 & 4.9$\times10^{68}$ & 7.3 $\pm$1.5 $\times$ $10^{39}$ & $-$0.37 & 2.6 &27.4 $\pm$1.1 $\times$ $10^{-14}$ & 9.1 $\times$ $10^{-14}$ & 57.8 \\
~77.3 & 6.8$\times10^{68}$ & 2.7 $\pm$0.5 $\times$ $10^{39}$ & $-$0.50 & 1.8 &12.3 $\pm$0.5 $\times$ $10^{-14}$ & 4.1 $\times$ $10^{-14}$ & 56.1 \\
115.9 & 7.4$\times10^{68}$ & 2.3 $\pm$0.4 $\times$ $10^{39}$ & $-$0.58 & 1.6 &10.7 $\pm$0.4 $\times$ $10^{-14}$ & 3.6 $\times$ $10^{-14}$ & 54.7 \\
154.5 & 1.4$\times10^{69}$ & 2.6 $\pm$0.5 $\times$ $10^{39}$ & $-$0.49 & 1.4 & 8.2 $\pm$0.3 $\times$ $10^{-14}$ & 2.7 $\times$ $10^{-14}$ & 52.7 \\
193.2 & 8.9$\times10^{68}$ & 1.7 $\pm$0.2 $\times$ $10^{39}$ & $-$0.60 & 1.5 & 8.5 $\pm$0.3 $\times$ $10^{-14}$ & 2.8 $\times$ $10^{-14}$ & 53.8 \\
231.8 & 5.7$\times10^{68}$ & 1.1 $\pm$0.2 $\times$ $10^{39}$ & $-$0.49 & 1.5 & 8.9 $\pm$0.4 $\times$ $10^{-14}$ & 3.0 $\times$ $10^{-14}$ & 53.8 \\
256.9 & 5.7$\times10^{68}$ & 1.2 $\pm$0.3 $\times$ $10^{39}$ & $-$0.67 & 1.5 & 8.8 $\pm$0.4 $\times$ $10^{-14}$ & 2.9 $\times$ $10^{-14}$ & 53.8 \\
\hline

\end{tabular}
\end{table*}

\subsubsection{IC\,310}

The core has a low spectral index value of $-$0.33. We
find that as one moves farther away from the core, the spectral index steepens,
consistent with earlier results by \cite{1998A&A...331..475F} and
\cite{1998A&A...331..901S}.
Unlike in PKS\,0053$-$016, the spectral index does
not flatten at the position of the flare.  This might be because the first
region at the core is dominated by the emission from the compact core instead
of the jet, and hence the difference in spectral index between the first and
second regions is barely noticeable. In PKS\,0053$-$016, the emission from the
compact core is not very prominent.
%The plot of the variation of the magnetic field along the tail (
Figure ~\ref{spec-in-eq-para} shows that there is a
discontinuity in the decreasing trend in the magnetic field seen farther
downstream. It is noticeable that the spectral index decreases at the location of this discontinuity.
The steady steepening of the spectral index despite
an increase in the magnetic field indicates that there are no reacceleration
processes playing a role in this region. Table~\ref{phy-para} for IC\,310 shows
that the net luminosity emitted is comparable to that in the other regions but from a
much smaller volume. Hence, this discontinuity in the magnetic field most
probably arises because of the projection effects that lead to uncertainties
while measuring the volume.  Toward the end of the tail, beyond 45 kpc,
we find a discontinuity in the trend of the spectral index, where it starts to
rise after a small dip and thereafter falls again. It is unlikely that this is
happening because of uncertainties in imaging the diffuse structure because, had it been
the case, the lower-frequency images will pick up more flux from diffuse
structures because of better $uv$ coverage and hence show an artificial
steepening trend instead of flattening.  

 \subsubsection{NGC\,1265}

The overall steepening of the spectral index toward the tail is perceivable in
both the spectral index map and the spectral index versus distance plot
(Figure~\ref{spec-in-eq-para}). This trend was previously observed by
\cite{1987A&AS...71..603J}, \cite{1998A&A...331..475F} and
\cite{1998A&A...331..901S} at higher frequencies. The spectral index varies
from $-$0.43 at the host galaxy to $-$0.8 at the tip of the detected tail. The
trend is quite smooth in both the magnetic field plot and the spectral index plot.
However, the spectral index map shows that the edges of the tails show a slightly
steeper spectral index than the ridge line. This might be an aftereffect of the reacceleration caused by the interaction between the two
separate tails at the center. This might also arise because of the systematic
deconvolution errors introduced while cleaning. 
The spectral index plot does not show a steepening trend after about 150 kpc.
However, the errors are quite high in this case, and hence we may not be
justified in favoring a flatter spectrum toward the end of the tail.

\subsubsection{GB6\,B0335$+$0955}

The spectral index map and its variation with the distance are shown in
Figure~\ref{spec-in-eq-para}.
%Toward the end of the tails, there are deviations from a steadily declining spectral index. However, these variations are within the error bars.
Though the intensity maps show a sudden
increase in intensity near the core, it is evident from the values in
Table~\ref{phy-para} that the core region is much more luminous than the second
brightness peak. %This brightness peak is not likely to be arising due to the
%typical flaring events usually seen in FR\,I radio galaxy. 
The brightness peak is
several tens of kiloparsecs away from the core. The spectral index is low at the core,
with a spectral index value of $-$0.2. The spectral index at the nearby, bright
blob-like feature is also quite flat, with a spectral index of about $-$0.26.
This is very unusual for a synchrotron-aged plasma unless it is being
reaccelerated by some other mechanisms. The slight change in direction of the
tails at the point raises suspicions that this is an aftereffect of
gravitational interaction with a neighboring galaxy.
%The spectrum steepens steadily beyond this region. 

\cite{1995ApJ...451..125S} also finds this steepening in the spectra. Their
spectral index values range from $-$1.27 to $-$2.02. Their spectral index values
are bigger than ours probably because the values were
calculated from high-frequency (between 1.4 and 4.88 GHz) maps, and hence a
break might have occurred. The total extent of the galaxy is also smaller
compared to our maps. The field is quite crowded, and the high-frequency images show
many other sources overlapping the tail. The steady steepening trend that is
usually seen in head$-$tail galaxies is not seen in this galaxy. Although there
is a rising trend in the spectral index, the net luminosity is decreasing
consistently along the tail.  
\subsubsection{IC\,711}

Figure~\ref{spec-in-eq-para} shows the spectral index map of IC\,711. The
spectral index varies from $-$0.7 at the core to $-$2.0 toward the end of the
tail. The discrepancy in the trend of the spectral index at about 350 kpc away
from the head was noted by \cite{1977A&A....58...79W}. They concluded this was
happening because of the acceleration of the particles or the injection of the energy
from any other external agency. But from our current understanding of radio
galaxies, it is extremely unlikely that an external source provides energy to
the tail. Though the spectrum is flatter compared to the nearby regions, the
net luminosity from this region follows a smooth decreasing trend. This implies
that, instead of an increase in the total energy, the flattening of the spectral
index is a result of the redistribution of energies of the electron population.
This might be induced by some turbulence due to the gravitational influence of
nearby massive galaxies like IC\,708, which appears to be very close to this
part of the tail. 

\subsubsection{NGC\,7385}

\cite{1975A&A....40..221S} have studied the variation of the spectral index
along the tail of this head$-$tail galaxy and have noted the progressive increase
in the spectral index, with the maximum value of the spectral index reaching up
to $-$1.0.
Figure~\ref{spec-in-eq-para-b} shows the
spectral index map and its variation with distance along with the variation of the
magnetic field with distance.
Note that NGC\,7385 is one of the galaxies in
the sample with a misplaced optical host. The second point in the plot hence
corresponds to the region which includes the core and is flatter compared to
the first and the third point. This also leads to a sudden increase in the
equipartition magnetic field. The signal-to-noise ratio was not high enough to
obtain the spectral index values of the extended structures with the low
surface brightness. The range of the spectral index is from $-$0.37 near the
core to $-$0.67 at the end.

\section{Discussion}
\label{ht-discuss}

Seven head$-$tail radio sources have been mapped in detail with an
angular-scale resolution of $\sim$5$^{\prime\prime}$--20$^{\prime\prime}$.
\subsection{General source properties}
\label{gsp}

%Below we discuss equipartition parameters and radiative ages
%of our sample sources.  We also discuss kinematics of some of the intriguing head$-$tail sources.
%Two important clues to understand the nature of head$-$tail radio galaxies are,
%(i) there could be a pattern of the morphologies of the tails about the
%cluster centers and (ii) the most massive, dominant galaxy is expected to have
%a smaller random velocity about the ICM and hence, 

All the head$-$tail galaxies have so far been found exclusively in clusters, and in most cases
in rich clusters.
Since the morphology of head$-$tail radio galaxies is thought to be a result of
the ram pressure that is due to the relative motion of the radio galaxy through the
ICM \citep{1972Natur.237..269M, 1998MNRAS.301..609B}, fitting the bending of the trajectory of jets using beam models
\citep{1979Natur.279..770B,1979ApJ...234..818J,1985A&A...143..136B}
gives estimates of several parameters, such as the jet flow velocity and the 
initial ejection angle of the jet with respect to the motion of the galaxy.
Hence, the hypotheses that host galaxy orbits are either radial, circular,
or isotropic are tested, and the hypothesis that head$-$tail
radio galaxies are primarily (or the inward portion of them) in highly
elliptical orbits \citep{1979ApJ...234..818J} is also tested.
%, which might be the case if such galaxies
%%required interstellar gas that was stripped during a passage
%through the cluster center and replenished when the galaxy was far from
%the center \citep{1979ApJ...234..818J}.
Observationally, images at higher angular resolution are needed to do such analyses.
Since most of the head$-$tail radio galaxies in the sample are
narrow-angle tail (NAT) sources, the two jets are just not
distinguishable, such as
in three of the seven sources in our sample.
Furthermore, the probability of projection effects to be playing a major role
is large, and it becomes difficult to decouple the projection effects and the
trajectory of the jets. %and clearly
% DONE by DVL
%In addition, an understanding of thermal particle densities inside the tail is
%needed, and hence high-resolution polarisation distributions of head$-$tail
%sources at several radio bands are also required.
In the light of our data for the sample sources and our understanding
from the literature, below we present arguments to state that
(1) the bending of radio tails is due to the motion
of the host galaxy through the ICM, (2) these galaxies have an isotropic
distribution of the orientations about the cluster center,
(3) the asymmetries seen in the intensity profiles of the radio jets
in some of the head$-$tail radio galaxies
are due to the difference in environments along with the projection effects
and the ejection of jets at acute angles with respect to
the direction of motion of the host galaxy through the ICM, and
(4) the multiple bends and wiggles are possibly due to the
precessing radio jets.

It was previously noted by \cite{1984ApJ...278...37E}, \cite{1982IAUS...97...45B}, and
\cite{1981MNRAS.195..523B} that the bending in radio jets seen in wide-angle tail (WAT) radio sources
could not be explained by the usual ram pressure bending models.  This is because these
galaxies are usually seen to be associated with the cluster dominant galaxies,
which would have low velocities against the ICM gas.
However, five out of the seven sources in our sample are part of rich, cool core
clusters and are seen to be at a range of distances from the brightest cluster
galaxy.  Although IC\,711 and NGC\,7385 are not part of rich clusters,
they are not associated with the dominant and the most massive galaxy of
their respective clusters.
We therefore expect that all of the galaxies in our sample would
have high velocities around their cluster center potentials.
As a result, the bending of the tails is entirely possibly due to the motion
of the host galaxy through the ICM. 

According to ram pressure bending models, the radio trails trace the motion of
the galaxy through the cluster. In a random sample of galaxies in a cluster,
one expects random directions of motion of galaxies.
We find that NGC\,1265 and IC\,310 are directed toward NGC\,1275, the cD
galaxy in the Perseus cluster. PKS\,B0053$-$016 is also almost directed toward
the cluster center. GB6\,B0335$+$0955 is an example of a galaxy directed away
from the cluster center, while the orbital motion of IC\,711 around its cluster
center is clearly visible due to its large size.  The orientation of the radio
tails of the rest of the sample head$-$tail radio galaxies seems to imply an
orbital motion of the host galaxy around the cluster center rather than a
radial infall.
\cite{1987ApJ...316..113O} using a sample of 70 NAT radio sources, concluded
that the head$-$tail sources are randomly oriented about the cluster center.
Although our sample is statistically
small, it seems that the random orientations of head$-$tail radio galaxies about the cluster
center are consistent with the notion that the distribution of the orientations
of these galaxies is isotropic
\citep{1976ApJ...203L.107R,1978MNRAS.183..195G, 1982IAUS...97...77H,
1987ApJ...316..113O}.
 
As was already mentioned in Sections~\ref{pksbmorph} and~\ref{ng7385}, two of
the five tailed galaxies with both jets separately visible in our 610 MHz high
resolution maps, the optical hosts are misplaced from the tip of the tails.
Additionally, the intensity profiles of the jets in IC\,711 are asymmetric
(see Figure~\ref{intensity-ridge} and discussion in
Sec.~\ref{roles-ism-icm}).
The northern jet in it is seen to flare out at a larger projected distance from the
core than the southern jet.
%This could be because of the orientation of the jet axis with respect to the motion of the galaxy.
This asymmetry in the jets in these three galaxies can occur if the jet axis is
not aligned at $90^{\circ}$ with the motion of the galaxy. However, this asymmetry
may be due to projection effects \citep{1980PASAu...4...74R}.
Since it is known that there is no known correlation between the optical axes of the
host galaxies and the jet axes of strong radio sources,
there is no reason to expect a correlation between the jet axis and
the motion of the galaxy in our sample of head$-$tail radio sources
\citep{1985ApJ...291...32B,2009MNRAS.399.1888B,2010ASPC..427..365B}.
Hence the asymmetry in the jets is probably either due to projection
effects or the ejection of the jet at acute angles
with respect to the direction of motion of the host galaxy through the ICM.
%acute angles with the direction of motion or a combination of both.

Four out of five head$-$tail radio galaxies show both
jets distinctly visible: namely, PKS\,B0053$-$016, PKS\,B0053$-$015,
GB6\,B0335$+$0955 and IC\,711.  They also show the presence of the multiple bends and wiggles,
which are likely due to the precessing radio jets. \cite{1981ApJ...246L..65I}
noticed similar features in the tail of 3C\,129 and fitted precession models to
the tail.  \cite{1987ApJ...316...95O} listed three possibilities for explaining
the wiggles that were seen in NGC\,1265: (1) precession, (2) helical
instabilities, and (iii) variations in the momentum flux that possibly caused these
wiggles. Though they do not rule out precession and variations in the momentum flux
completely, they prefer helical instabilities over the other two possibilities.
PKS\,B0053$-$016 has a symmetric structure.
Though the helical features are not
completely symmetric in the other three galaxies, it is noticeable that every
loop or wiggle in one jet has a counterpart in the other. The other asymmetries possibly arise from the above-mentioned effects, namely
acute angles between
the direction of motion and jet axis, projection effects, and differences in
their environments.
Although NGC\,7385 does not show any helical structures at the length scales
seen in the rest of the four sources, the low-resolution images of
it (Figure~\ref{n7385}) show bends toward the ends of the tails, and it is not
possible to completely rule out the possibility that the origin of these bends
is similar to the ones seen in the rest of the four sources.
For example, if
precession leads to these features, then a smaller precessional velocity
than the advancing speed of the galaxy through the ICM \citep{1998MNRAS.301..609B}, might explain
these large-scale bends.
%However, the frequent occurrence of such structures in tailed sources yet need to be explained.

\subsection{Spectral Structure}

We list a few qualitative
points on the overall source spectral structures,
based on the images presented in Figures~\ref{all-ht}--\ref{spec-in-eq-para} above.
\begin{itemize}
\item[(i)] A comparison of 610 and 240/325 MHz maps shows that large variations of spectral index ($\Delta\alpha \simeq$2.2) occur in the tails of all head$-$tail radio galaxies.  
\item[(ii)] On moving along the tail from the head (or optical host galaxy), the spectrum typically steepens toward the ends of the tails. 
\item[(iii)]In the region where flaring occurs, there is a net increase in the total luminosity along with the flattening. This suggests that the particles are re-accelerated in or near this region. 
\item[(iv)] A steady decline in spectral index along the tail is not always seen in some of the galaxies. However, in most of these cases, the equipartition magnetic field and the net luminosity at the position of the discrepancy continue to follow the declining trend. Hence these discontinuities in the general steepening trends may be due to some turbulence that is present in the cluster environment, which leads to redistribution of the the energies of the electron population. It is possible that this turbulence is caused by the galactic wakes of the nearby galaxies \citep{1979ApJ...234..818J}, by their gravitational influence.
\item[(v)] The radiative age estimates are consistent with the hypothesis that the radio
plasma ages due to synchrotron cooling as it moves away from
the head of the radio galaxy, and the radio plasma at the farthest end
is the oldest.
\item[(vi)] The equipartition magnetic field and also the pressure in the farther
end of the tail is lower by $\sim$4 as compared to the region close to the core,
which is similar to the result for 3C\,129 \citep{2004A&A...420..491L}.
%The equipartition magnetic field in the system can be treated as 
%a proxy for the net energy in the system.
There is a clear decrease in the spectral index and in the magnetic field values as one moves away from the core.
%The assumption that there is equipartition of energy between the radiating particles and magnetic field seems to break at certain regions in several sources, and the estimated radiative age peaks at these regions as compared to the adjacent regions.
%The spectral index is also plotted against the distance from the core.
\item[(vii)] In several galaxies, there is a jump in the steepening of the
spectrum, such as in IC\,711, where the jump in spectral index toward the end is quite
significant.
Similar complexities are also seen in IC\,310, NGC\,7385, NGC\,1265 and GB6\,B0335$+$0955.
% although the deviations in trends is not significant in all of them.
Such spectral index jumps were noted earlier by \cite{1977A&A....58...79W} as well,
and this suggests reacceleration of electron population \citep{1973A&A....26..423J,2004A&A...420..491L} in these regions 
due to shocks or turbulence in the cluster environment.

It is interesting to note the presence of IC\,708,
another massive head$-$tail radio galaxy close to this region showing a jump in spectral index. %and
It is possible that this region of IC\,711 showing a jump in spectral index is also under the
gravitational influence of IC\,708 
(Figure~\ref{spec-in-eq-para}). The interaction with nearby galaxies is also seen in GB6\,B0335$+$0955. Hence the turbulence in the ICM induced by the random motions of galaxies passing close by could lead to the reacceleration of the particles in the tail.
\item[(viii)] The energy of synchrotron-emitting electrons is redistributed in regions showing
discontinuities in the spectra, i.e., the lower-energy electrons
are energized to higher energies and hence the spectral index flattens.
Furthermore, the total energy in these particles is conserved; in other
words, since the flux density is seen to be reduced compared to the
neighboring regions, there is no source of a fresh injection of electrons.
%As suggested by \cite{}, particle acceleration may also play
%a role as seen in the tails of 3C\,129  in order to
%explain the differing behavior of spectral index.
%
% THIS GETS REPEATES (DVL)
%This trend is consistent with our expectations as was discussed in Section ~\ref{specstr}.
%Briefly, the  electron population in the tails lose
%a significant amount of its energy by radiation.
%If the particles are in equipartition,
%then the total energy will be shared almost equally as magnetic field energy
%and electron energy.
%Hence the equipartition magnetic field becomes weaker with distance from the head.

%It is, therefore, possible that the particle acceleration is required in the
%tails of IC\,310 and IC\,711.
\end{itemize}

\subsection{Dynamical age}

The head$-$tail radio structure is strongly suggestive of the following scenario:
the radio galaxy, in its orbital motion in the cluster, moves rapidly through the ICM;
its radio components are slowed down by the ICM and lag behind the radio galaxy; and the farthest part of the radio source lags behind the radio galaxy.
If this picture is correct, the velocity of the radio galaxy through the ICM
and the radio properties of the head$-$tail sources give information on the time of ejection
of relativistic particles by the galaxy and on the density of the ICM.
Therefore, from the difference in cluster redshift and the galaxy redshift,
the line-of-sight velocity of the galaxies was estimated.
%The redshift values were obtained from NED.
Further, assuming equal velocities along all three axes, the 
velocity in the plane of the sky would be on average $\surd2$ times the line-of-sight velocity.
This and the projected length of the tails,
which is used as a proxy for the distance traveled by the host galaxy,
were used in determining the dynamical ages of the head$-$tail radio galaxies.
\begin{table}[tbph]
        \centering
        \caption{Dynamical age estimates of our sample sources.}
        \label{dyn-age}
 \begin{tabular}{lrrcr}
\hline
\hline
Object       &  Velocity  &  \multicolumn{2}{c}{size} & age \\
          &  (km~s$^{-1}$ & ($\prime$) & (kpc) & (Myr) \\
\hline
PKS\,B0053$-$016&   785.1 &  5.3 &  238.9 &  295 \\
PKS\,B0053$-$015&  3105.7 &  7.2 &  330.7 &  105 \\
IC\,310         &   540.0 &  7.5 &  165.3 &  300 \\
NGC\,1265       &  3760.5 & 10.0 &  220.4 &   55 \\
GB6\,B0335$+$096&   961.3 &  8.2 &  358.9 &  365 \\
IC\,711         &   552.9 & 17.8 &  720.6 & 1275 \\
NGC\,7385       &   428.2 & 14.3 &  442.3 & 1010 \\
\hline
 \end{tabular}
 \end{table}

Table~\ref{dyn-age} presents these dynamical age estimates for our sample
head$-$tail radio sources.
Note that any head$-$tail radio galaxies which are nearly aligned along
the line of sight may not be recognized as head$-$tail sources.
These head$-$tail radio galaxies have relatively large velocities with respect to the ICM
because of their motion along the line of sight. 

A spectral aging analysis was done for the head$-$tail galaxies in Perseus cluster
by \cite{1998A&A...331..901S}. The spectral age of NGC\,1265 from their
analysis and our dynamical ages are within the same order of magnitude.  But
the dynamical ages of IC\,711 and NGC\,7385 are about an order of magnitude higher,
and that of NGC\,1265 is an order of magnitude lower
than their spectral or the radiative ages,
whereas for the rest of the sample sources, the spectral ages and the dynamical
ages are of the same order. This
could mean that the direction of motion of NGC\,1265
is not aligned to our line-of-sight whereas the direction of motion of IC\,711 and NGC\,7385 are aligned toward
our line of sight.  Given that the sample consists of large
known head$-$tail radio galaxies, it is highly probable that the plane of the sky
velocity is higher than what is assumed in most cases. Hence it is
likely that the dynamical ages, presented in Table.~\ref{dyn-age} are an
underestimation for NGC\,1265 and an overestimation for IC\,711 and NGC\,7385.
The dynamical ages of all of the galaxies in our sample are (0.6--12.8)~$\times 10^8$ yr, with the median being 3.0~$\times 10^8$ yr,
are a factor of 1--20 more than the radiative age estimates of
(0.4--0.6)~$\times 10^8$ yr, with the median being $\sim$0.5~$\times 10^8$ yr for our head$-$tail sample sources.

\subsection{Roles of interstellar and intracluster media}
\label{roles-ism-icm}

\begin{figure*}
\begin{center}
\begin{tabular}{c}
\includegraphics[width=18cm]{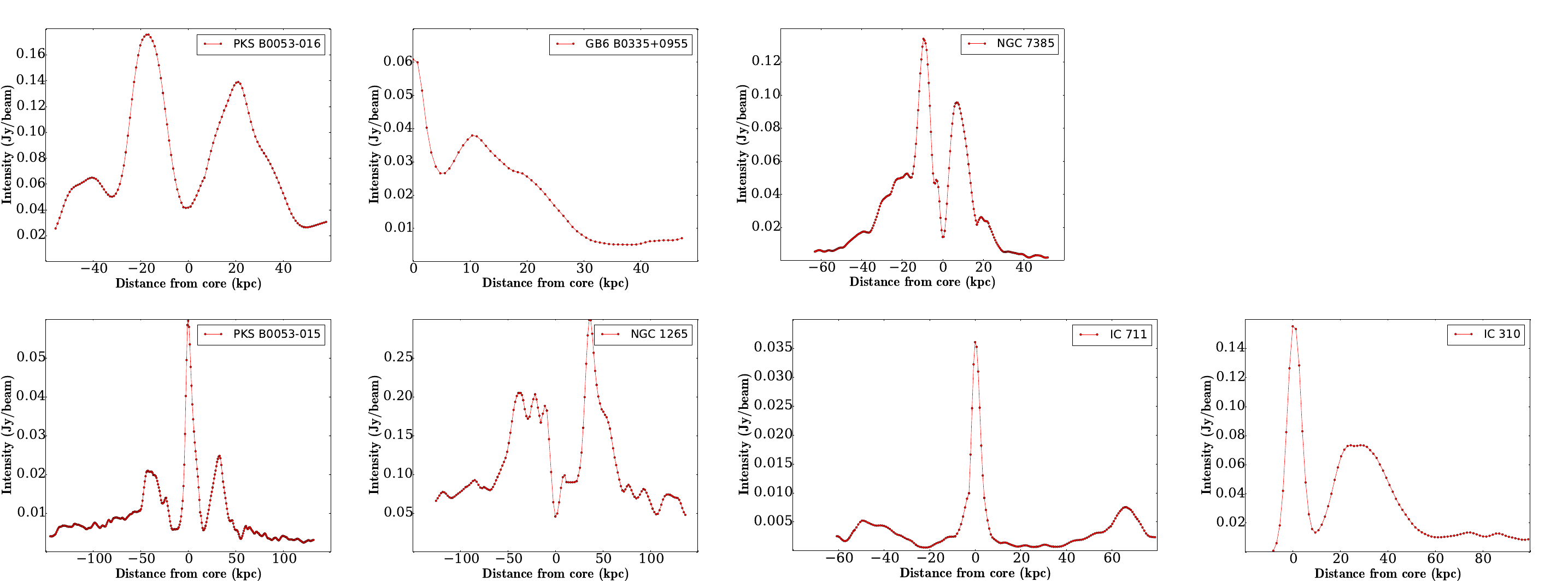}
\end{tabular} 
  \caption{The spatial profile of radio intensity
in the inner parts at 610 MHz as a function of distance from
radio core along the ridge line.
The profiles for all our sample sources are ordered as, clockwise from top
left,PKS\,0053$-$016, GB6\,B0335$+$0955, NGC\,7385, IC\,310, IC\,711, NGC\,1265
and PKS\,0053$-$015, same as Fig~\ref{all-ht}.
The radio intensity profiles are shown along both radio jets, except
for GB6\,B0335$+$0955 and IC\,310.
The position of the peak radio intensity away from the
radio core along the ridge line marks the location of the flaring.}
  \label{intensity-ridge}
\end{center}
\end{figure*}

%\cite{1979ApJ...234..818J} noticed that the flaring up of the beams into
%plumes occur at scales similar to the extent of ISM which will be retained
%within an elliptical galaxy moving through the ICM. They take into account the
%effects of the ram pressure stripping of the ISM which then is compensated by
%the mass loss rate from the stellar winds. 

\citet{1972ApJ...176....1G} proposed that the
infall of galaxies onto cluster potential during the early
phases leads to ram pressure stripping of the ISM in these galaxies.
%little or no intergalactic gas.
Hence, elliptical galaxies, especially those in clusters, are poor in
gas content.
However, \cite{1979ApJ...234..818J} show that elliptical galaxies moving
through a cluster can still retain a significant amount of
ISM within $\sim$10~kpc, called as the influence of the ISM.
%in such galaxies is stripped
%by ram pressure stripping \citep{1972ApJ...176....1G}. However,
%\cite{1979ApJ...234..818J} show that a significant amount of ISM can still be
%retained within the inner regions ($~$10~kpc) of these galaxies.
Since in some radio sources
the scales of bending usually seen in head$-$tail radio sources
are similar to the size of influence of the ISM, it suggests that the ISM
may be important in the formation of head$-$tail radio sources.
Qualitatively, when a galaxy is moving at a transonic
%or mildly relativistic
speed through the ICM, a bow shock is expected to develop at the leading edge
of the galaxy, assuming that some amount of ISM is retained within the galaxy
\citep{1979ApJ...234..818J}. Consequently, a turbulent galactic wake would be
formed behind the galaxy, %and that the bending
%could also take place within the ISM due to the headward pressure gradient
%within the ISM 
%The radio jets
%are deflected into the galactic wake in which case the net luminosity of the
%tails couldn't be explained using the kinetic energy of the beams. Hence, they
%invoked the tapping of energy from the turbulent wake behind the galaxy to
%explain the reacceleration of particles in the beams.
and that the flaring seen in head$-$tail galaxies occurs when the radio jets are
re-accelerated as a result of the turbulence in the galactic wake \citep{1979ApJ...234..818J}.
Hence, in this context, below
we discuss if the head$-$tail radio galaxies in general show
similar deceleration phenomena and at what size scales.

Head$-$tail radio sources are considered
to form a subset of the FR\,I type sources and all FR\,I sources, have a region
with a sudden increase in brightness close to the core.
The initially relativistic jet is thought to be decelerated to sub-relativistic
speeds at the flaring point, which is due to
entrainment of the surrounding material or by injection of stellar ejecta \citep{1994ApJ...422..542B, 1994MNRAS.269..394K, 1995ApJS..101...29B}.
The flux ratio of the jet to its counterjet along the distance from the core suggests
that an initially relativistic jet with velocity  $\sim0.6c$ decelerates to
sub-relativistic speeds, $\sim0.1c$ within $\sim$5~kpc \citep{1994ASPC...54..241P}.
%
%We next look at the size scales at which flaring of the jet happens in our
%sample of head$-$tail radio galaxies.
Figure ~\ref{intensity-ridge} shows plots of radio intensity in the inner parts
at 610 MHz as a function of distance from the core along
the ridge line for our sample of head$-$tail radio sources.
%The low-resolution maps of NGC\,1275, IC\,310 \& PKS\,B0053-016
%was used because of the presence of too many substructures or deconvolution
%errors.
%The spatial profile of NGC\,1265 is complicated due to the presence of
%many knots.
Although the flaring is not always very smooth in every galaxy, 
the projected lengths, on an average, are always $\gtrsim$5~kpc, whereas,
according to \citet{1979ApJ...234..818J}, the location of flaring must depend on the size
scale of the ISM, which depends on the ICM density \citep{1989A&A...225..333G}, ram pressure and so on.
The location of the maxima of the flaring varies from
$\simeq$15 kpc in NGC\,7385 and GB6\,B0335$+$0955 to
$\gtrsim$60 kpc in IC\,711.
In addition, there are some asymmetries at the location of the flare in
several head$-$tail radio galaxies, such as in IC\,711, which
might be due to several reasons discussed in Section~\ref{gsp}.
These differences in the distances of the flaring in our sample sources could
be attributed due to the difference in environments along
with projection effects and ejection of the jet at acute angles
with respect to the direction of motion of the host galaxy through the ICM.
Additionally, an independent study of WATs \citet{ODonoghue}
found a range from $\sim$15 to $\sim$119~kpc for locations
of the radio flaring.  
It is also possible that the distances at which flaring occurs
in our sample of head$-$tail radio sources may appear small due to projection, and hence actual sizes of the flaring region may be larger.
Nevertheless, it is clear that flaring can appear at locations as large as
$\sim$60~kpc, in our sample or $\sim$119~kpc in \citet{ODonoghue} sample sources,
and it is unlikely that the ISM extends to such large values.

%Although WATs are usually associated with the cD galaxy of a cluster and hence might retain a significant amount of ISM their velocities might not be large enough to produce a bow shock.
Instead, flaring at such large distances from the core are likely due to environmental disturbances,
such as from the entrainment by hot ICM, consistent with \cite{1994ApJ...422..542B}, \cite{1994MNRAS.269..394K} and \cite{1995ApJS..101...29B}.
Indeed, the ICM within merging clusters of galaxies is likely
to be in a violent or turbulent dynamical state and may have significant
effect on the evolution of cluster radio sources \citep{1998MNRAS.301..609B}.
The three-dimensional simulations of jets propagating in such ICMs
show that a collimated radio jet encountering a
surface brightness edge in a merging cluster environment disrupts and flares
as a result of Kelvin--Helmholtz instabilities \citep{1995ApJ...445...80L}.
%
%From table ~\ref{phy-para} it is evident .
%
%PKS\,B0053-016 is an example in which core is not very dominant, and the
%spectrum flattens in the flaring region.  The core to flaring point distance
%was previously noted to reach a maximum value of 10 kpc in the B2 sample by
%\cite {1999MNRAS.306..513L} .
%Given that the flaring is very similar except for the scales at which these
%occur, the basic physical origin might be similar. It was proposed previously
%that the difference in FR-I and FR-II radio galaxies might be the intrinsic
%strength of the host galaxy and the environment.
%%Probably this explains why most head$-$tail sources are FR-I type morphology. 

\section{Summary}
\label{ht-concl}

We have presented the GMRT data for the seven large known head$-$tail
radio galaxies.
By examining the gross characteristics,
several conclusions are drawn by examining the gross morphological
structures,
spectral structures, equipartition parameters, and kinematics
of these head$-$tail radio galaxies.  
\begin{enumerate}
%\item All sample head$-$tail radio sources are consistently described
%in terms of relative motions of radio galaxies through the ICM
%in clusters of galaxies.
%
%\item Head$-$tail radio sources are not associated with the single dominant
%galaxy in a rich cluster.  This is consistent with the hypothesis that
%motion through the ICM is responsible for the tails.%, and that
%single dominant galaxies are at rest with respect to the ICM.

%\item  

\item The sample is composed of seven large head$-$tail galaxies in terms of their
angular sizes. All these were found to be hosted by prominent and massive
elliptical galaxies. 
Additionally, massive neighboring galaxies can distort a simple orbital motion
trajectory of head$-$tail radio galaxies, as is seen in
GB6\,B0335$+$0955, IC\,711 and NGC\,7385.

\item Five sources in the sample show clear, distinguishable radio
tails or jets seen in high-resolution images, two sources have their optical hosts away from the tip of
their radio emission tails.
This suggests that there is no preferential orientation of
the radio jet with respect to the direction of motion of the host galaxy. 

\item Four of the five head$-$tail radio galaxies,
namely, PKS\,B0053$-$016, PKS\,B0053$-$015, GB6\,B0335$+$0955 and IC\,711
with clear distinguishable radio tails/jets show the
clear presence of the multiple bends and wiggles.
The presence of these wiggles is possibly due to the
precession of the jet.

%\item NGC\,7385 does not show wiggles,
%but instead, shows the multiple bends, which is
%suggestive of a change in the direction of radio jet.
%It is possibly due to
%change in the direction of motion of the host galaxy or
%it is just due to the projection effects.  Hence, it is highly
%probable that the angle of the motion of the host galaxy to our line of sight
%keeps changing throughout its motion.

\item In almost all of the sources in the sample, we see a flaring in the
intensity of the jet close to the core; the radio jet spreads out
and disappears into lower surface brightness radio emission.
%It is possible that
%at the location where the jet flares, the jet begins to propagate into
%regions of lower surface brightness beyond the flare.
The spectral index of the radio jet is flatter near the core.
It begins to flare as it encounters a surface brightness edge, and the spectral index
begins to steepen immediately after the location of flaring.  The spectral index
is the steepest in regions with lower surface brightness,that is, at the farthest end of tails. 
%This flaring is quite common in FR\,I
%radio galaxies and \cite{2002MNRAS.336.1161L} explained this flaring as a result
%of the entrainment of the jet material by the external environment leading to
%its deceleration from relativistic to sub-relativistic speeds.
Furthermore, the bulk kinetic
energy of the jet is probably being converted into the internal energy,
which explains the possible increase in the total radio emission along
with a relatively flat radio spectrum. 

\item Since projection effects result in the classification of many
of the WAT sources as NAT sources,
%with wide angles of ejection of jets as head$-$tail sources,
these NAT sources are our best bet to look directly into a spatially extended
radio source nucleus according to unification theory, as evidenced in the case of IC\,310.
%For example, the classification of the IC\,310 as high-energy peaked BL~Lac
%(\cite{2014A&A...563A..91A}).
%indeed indicates that the inclination angle of this host galaxy
%with respect to our line of sight is small. 

\item By examining the gross spectral structure,
we find steepening of the radio spectrum in all head$-$tail radio galaxies,
and we do not see evidence for a region with a constant spectral index
(see Section~\ref{intro}).  We believe this is due to
coarser sampling of regions, and within error bars, the results are consistent.
%But visually the observed spectral index is mostly continuously decreasing.

\item The equipartition magnetic field is also seen to fall as distance from the head along the tail increases.
But there are regions where the magnetic field does not show a smooth decline,
whereas the spectral index shows a smooth decline.
Since we assumed cylindrical symmetry,
the equipartition magnetic field is proportional to $(L/V)^{2/7}$
and given that the radio tails might overlap at least in projection,
we might be underestimating the volume at such regions, which in turn leads
to overestimation of the equipartition magnetic field.

\end{enumerate}

In a subsequent paper, we will discuss the radio
intensity and surface brightness gradients and the X-ray properties of these head$-$tail
radio sources.  Furthermore, since the X-ray data are indicative of an ICM,
there may be a contact surface brightness edge just before the
flaring of the radio jet, as suggested by \cite{1995ApJ...445...80L}.
Therefore, we will also investigate the morphological properties of a
propagating jet, say as it crosses its ISM into an ICM.
Also, a comparison of the particle density inside the tails with respect to
the external gas density via thermal bremsstrahlung would be investigated
in order to probe the conditions applicable for buoyancy or falling
in this head$-$tail radio galaxies sample.

\section*{Acknowledgments}

We thank the anonymous referee for his or her prompt review of this
manuscript and for comments that led to the improvement of this paper.
We also thank D.~A. Green, P. Kharb and S.~N. Murthy, and S.~Vaddi for discussions and several useful comments.
We thank the staff of the GMRT who have made these observations
possible. The GMRT is run by the National Center for Radio Astrophysics
of the Tata Institute of Fundamental Research.
This research has made use of the NASA/IPAC Extragalactic Database (NED), which is
operated by the Jet Propulsion Laboratory, California Institute of Technology, under contract
with the NASA, and NASA's Astrophysics Data System.

%%\noindent
%%{\bf {References}}
%\nocite{*}
%%\bibliographystyle{unsrt}
%\bibliography{refpropo.bib}

\clearpage

\end{document}